\documentclass[prb,twocolumn,floatfix]{revtex4-1}
\usepackage{graphicx,color,epsfig,rotate}
\usepackage{amssymb,amsmath,bm}  
\usepackage{dcolumn}  
\usepackage{hyperref} 
\usepackage[usenames,dvipsnames]{xcolor}
\usepackage{subfigure}
\usepackage{verbatim}

\newcommand{\g} \textbf{}

\def\smb{SmB$_6$}

\begin{document}

\title{In-gap collective mode spectrum of the Topological Kondo Insulator SmB$_6$}

\author{W. T. Fuhrman}
\affiliation{Institute for Quantum Matter at Johns Hopkins University, Baltimore, MD 21218, USA}

\author{P. Nikoli\'c}
\affiliation{School of Physics, Astronomy and Computational Sciences,\\George Mason University, Fairfax, VA 22030, USA}
\affiliation{Institute for Quantum Matter at Johns Hopkins University, Baltimore, MD 21218, USA}

\date{\today}

\begin{abstract}

Samarium hexaboride (SmB$_6$) is the first strongly correlated material with a recognized non-trivial band-structure topology. Its electron correlations are seen by inelastic neutron scattering as a coherent collective excitation at the energy of 14 meV. Here we calculate the spectrum of this mode using a perturbative slave boson method. Our starting point is the recently constructed Anderson model that properly captures the band-structure topology of SmB$_6$. Most self-consistent renormalization effects are captured by a few phenomenological parameters whose values are fitted to match the calculated and experimentally measured mode spectrum in the first Brillouin zone. A simple band-structure of low-energy quasiparticles in SmB$_6$ is also modeled through this fitting procedure, because the important renormalization effects due to Coulomb interactions are hard to calculate by ab-initio methods. Despite involving uncontrolled approximations, the slave boson calculation is capable of producing a fairly good quantitative match of the energy spectrum, and a qualitative match of the spectral weight throughout the first Brillouin zone. We find that the ``fitted'' band-structure required for this match indeed puts SmB$_6$ in the class of strong topological insulators. Our analysis thus provides a detailed physical picture of how the SmB$_6$ band topology arises from strong electron interactions, and paints the collective mode as magnetically active exciton.

\end{abstract}

\maketitle

\section{Introduction}\label{secIntro}

Kondo insulators are ``heavy fermion'' materials whose quasiparticle excitations have band-insulating dynamics \cite{Menth1969, Nickerson1971, Hundley1990, Riseborough1992, Alekseev1993, Nyhus1995, Sera1996, Okamura1998, Bouvet1998, Gorshunov1999, Riseborough2000}. The most studied Kondo insulator is samarium hexaboride (SmB$_6$), but other similar materials are also known (e.g. YbB$_{12}$, Ce$_3$Bi$_4$Pt$_3$, CeNiSn, CeRhSb, Ce$_3$Pt$_3$Sb$_3$, UNiSn). In the simplest physical picture of Kondo insulators, the quasiparticles are electrons from the hybridized atomic $d$ and $f$ orbitals of the material's rare earth element. The hybridization caused by the crystal environment opens a band-gap in the quasiparticle spectrum, which can be renormalized by interactions to a small value, about $19\textrm{ meV}$ in SmB$_6$ \cite{PhysRevB.59.1808}.

The history of SmB$_6$ goes back to at least 1969. Numerous experiments have revealed strong effects of Coulomb interactions among the samarium's ``heavy'' $f$ electrons. The quasiparticle spectrum of SmB$_6$ is significantly renormalized by interactions in a temperature-dependent manner \cite{Hundley1994}. Even more strikingly, inelastic neutron scattering (INS)  \cite{Alekseev1993, Kasuya1994, Bouvet1998, Fuhrman2014} has identified a coherent collective mode in SmB$_6$. The energy of this mode is smaller than the quasiparticle excitation gap (charge gap), and thus protects it from decay. Given that the lowest energy excitation is a bosonic mode, SmB$_6$ can be considered a strongly correlated insulator. Still, the $f$ electrons are likely not localized by Coulomb interactions in SmB$_6$, judging by the observed mixed valence of samarium \cite{Chazalviel1976, Beaurepaire1990}. This supports the picture of ``dressed'' hybridized electrons as quasiparticles, whose scattering in the particle-hole channel produces a collective exciton mode.

The interest in SmB$_6$ and other Kondo insulators has been renewed recently by the realization that their band-structure may have non-trivial topology \cite{Dzero2010, Takimoto2011, Dzero2012, Alex2013, Dzero2013}. As a putative strong topological insulator (TI), SmB$_6$ is expected to have a metallic crystal boundary protected against any source of backscattering that respects the time-reversal (TR) symmetry. Indeed, several transport \cite{Zhang2013, Wolgast2013, Kim2013, Kim2013a, Thomas2013, Phelan2014}, quantum oscillation \cite{Xiang2013} and surface spectroscopy \cite{Rossler2013, Denlinger2013, Neupane2013, Jiang2013, PhysRevB.88.121102} experiments have directly probed this metallic boundary and generally provided evidence for its existance. At the same time, SmB$_6$ has been heralded as the first true TI with a fully insulating bulk \cite{PhysRevB.88.180405}, unlike the original bismuth-based TIs whose bulks remain weakly conducting due to statistically unavoidable crystal defects \cite{Zhang-2009zr, Xia:2009, Hsieh:2008}. More importantly, SmB$_6$ is the first strongly-correlated TI as a member of the heavy fermion family of materials. Therefore, its metallic boundary is likely susceptible to various forms of strong electron correlations \cite{Roy2014, Nikolic2014b}.

In this paper we focus on the bulk collective mode of SmB$_6$ as a direct indicator of strong correlations and an indirect indicator of band-structure topology. We present the detailed theoretical analysis of the mode's dispersion that was able to reproduce the recent INS measurement with a reasonable quantitative accuracy \cite{Fuhrman2014}.

Inelastic magnetic neutron scattering provides direct access to spatial and temporal information about spin fluctuations. INS measurements were previously reported for both polycrystalline and single crystal SmB$_6$. Comprehensive investigations of phonons in SmB$_6$ have mapped out all acoustic, and the lowest optical phonon branches in the main symmetry directions \cite{SmB6LatticeDynamics}. Further INS measurements characterized magnetic excitations in single crystalline SmB$_6$. Two types of excitations were tentatively identified as (1) high energy ($40 \textrm{ meV}$ and $130 \textrm{ meV}$) intra-$J$-multiplet excitations of the short lived integral-valence 4f$^6$ and 4f$^5$ states, and (2) a coherent bound state at approximately $14 \textrm{ meV}$. The intensity of the coherent $14 \textrm{ meV}$ mode at the wavevector ${\bf Q} = (0.5,0.5,0.5)$ observed by INS rapidly increases as the sample is cooled below the insulating transition at $30 \textrm{ K}$, suggesting that it originates from hybridization \cite{Miyazaki2012}. Such properties of the magnetic excitation spectrum are intimately connected with the nature of the mixed valent ground state wavefunction.

Our calculation of the $14 \textrm{ meV}$ collective mode spectrum applies the perturbative slave boson method to the variant of the Anderson lattice model that is capable of capturing a non-trivial band topology \cite{Dzero2010, Dzero2012}. Several approximations are introduced to enable numerical calculation of the collective mode dispersion from a detailed modeled quasiparticle band-structure without jeopardizing any important renormalization effects. In this manner, we are able to treat the quasiparticle spectrum as a multi-component variational parameter. Ideally, the band-structure parameters are to be varied until the best match is found between the calculated and experimentally measured mode dispersions in the entire first Brillouin zone. This procedure thus provides indirect means to extract band-structure information from the INS data. It can have only a limited quantitative accuracy, but sufficient for the reliable determination of the band topology. As an experiment-based method, it can complement ab-initio calculations \cite{Kang2013, Lu2013b, Werner2013, Legner2014} whose accuracy in the low-energy sector is significantly limited by the presence of strong interactions.

So far, we considered a set of simple tight-binding band-structures on the cubic lattice of samarium atoms with up to third-neighbor hopping in both $d$ and $f$ orbitals, instead of freely varying the band-structure parameters. Some of these band-structures were inspired by ab-initio results. Remarkably, however, the best model we found was simple, with dominant third-neighbor hopping at least in the $d$ orbital. Consistent with this emerging picture are not only the ARPES findings of deep pockets surrounding X points \cite{Denlinger2013, Neupane2013, Jiang2013}, but also the chemistry of SmB$_6$.  Lying within the Bravais lattice of samarium atoms, the B$_6$ cluster has molecular orbitals lying along the main body diagonal in a t$_{1u}$ state.  This orbital is the LUMO state of the cluster when samarium is in the Sm$^{2+}$ configuration and facilitates third-neighbor exchange.  Furthermore, INS data from our previous work supports such a model, and the topology of the underlying band structure established SmB$_6$ as a strong TI \cite{Fuhrman2014}.

Our perturbative calculation paints a plausible physical picture of the collective mode as an exciton with internal ``multiplet'' states that transform non-trivially under TR. Similar theoretical descriptions have been proposed in the past, both from the intermediate-valence \cite{KikoinIV} and localized moment \cite{Riseborough1992, Riseborough2000, Riseborough2001} points of view. We improve on these studies in multiple ways. First, our calculations are based on the Anderson model of SmB$_6$, which is appropriate for the experimentally observed intermediate-valence dynamics. This Anderson model is further crafted to describe the realistic $d$-$f$ hybridization due to the crystal fields and spin-orbit coupling. Spin-orbit coupling is essential for the non-trivial topology, but has not been considered in the earlier collective mode studies. Second, we implement a perturbative RPA  calculation at the level of a two-body correlation function. This captures the emergence of an exciton bound state. Earlier RPA studies focused on self-energy corrections to the unbound particle-hole susceptibility, brought by the spin-exchange. We find that such self-energy corrections flatten the dispersion of the bound state. Lastly, our method can take a realistic quasiparticle spectrum and attempt to quantitatively describe an experiment while requiring only moderate numerical treatment.

This paper is organized as follows. Section \ref{secPrelim} describes the minimal model of SmB$_6$ and the slave-boson method. Section \ref{secSB} presents the perturbative collective mode calculation that utilizes the slave boson method, and explains both the analytical and numerical steps. Section \ref{secExp} compares the theoretical calculation with the neutron scattering data and draws the essential conclusions about both the quasiparticle and collective mode spectrum. All findings of the paper are summarized in the final section \ref{secConclusions}.

\section{Preliminaries}\label{secPrelim}

\subsection{The minimal model}\label{secModel}

\begin{figure}
\subfigure[{}]{\includegraphics[height=1.2in]{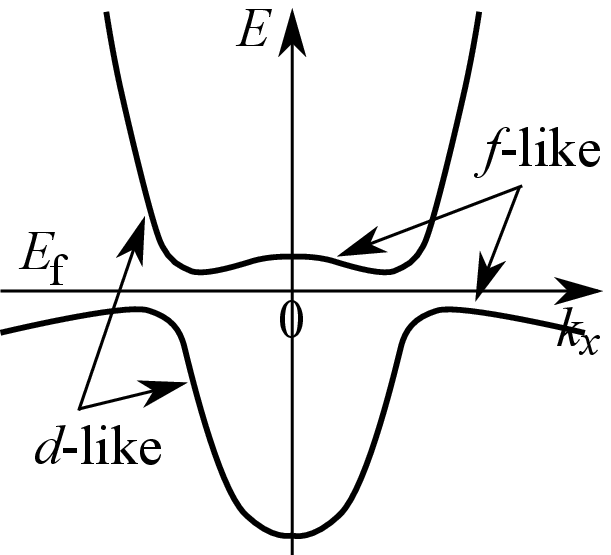}}\hspace{0.2in}
\subfigure[{}]{\includegraphics[height=1.2in]{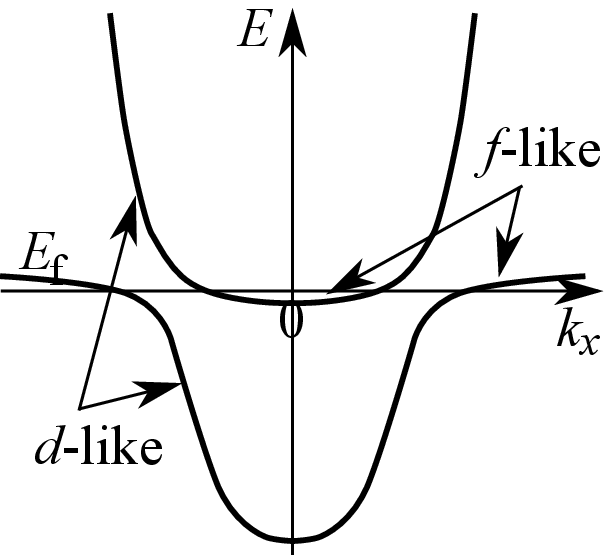}}
\caption{\label{HybGap}(a) A schematic band-structure of SmB$_6$. (b) Hybridization does not produce a bandgap without inverted band dispersions.  Note that the formation of an insulating state upon hybridization, as in (a), forms extrema in the region of the Fermi-level.  These extrema define a gapped ``pseudo Fermi surface'' and play the prominent role in the low energy dynamics of the spin-exciton.}
\end{figure}

The relevant degrees of freedom of SmB$_6$ are electrons that originate from the $d$ and $f$ orbitals of samarium atoms. The cubic crystal fields split the original sextuplet of the samarium's $f$ orbitals into a doublet and quadruplet, while hybridizing them with the $d$ orbitals due to the lost rotational symmetry at samarium sites. The resulting band dominated by the $d$ orbitals has a broad energy dispersion, while the bands dominated by the $f$ orbitals are nearly flat and have a large effective mass. In SmB$_6$ and all Kondo insulators, the intrinsic $d$ and $f$ bands have inverted dispersions at all momenta where they hybridize. This is a prerequisite for a bandgap to open as a result of hybridization, as shown in Fig.\ref{HybGap}. A Kondo insulator such as SmB$_6$ forms because its Fermi level sits inside this hybridization gap.

The essential dynamics of SmB$_6$ is described by the following tight-binding Hamiltonian defined on the cubic lattice of Sm atoms \cite{Dzero2010, Dzero2012}:
\begin{eqnarray}\label{MinMod}
H &=& \sum_{l}\sum_{\sigma}\int\limits_{\textrm{1BZ}}\frac{d^{3}k}{(2\pi)^{3}}\,
      \xi_{l{\bf k}}^{\phantom{\dagger}}d_{l\sigma{\bf k}}^{\dagger}d_{l\sigma{\bf k}}^{\phantom{\dagger}}\\
&& +\sum_{\alpha}\int\limits_{\textrm{1BZ}}\frac{d^{3}k}{(2\pi)^{3}}\,
      \epsilon_{\alpha{\bf k}}^{\phantom{\dagger}}\nonumber f_{\alpha{\bf k}}^{\dagger}f_{\alpha{\bf k}}^{\phantom{\dagger}}\\ 
&& +\sum_{l}\sum_{\alpha\sigma}\sum_{{\bf R}{\bf R}'}\Bigl(V_{l\sigma\alpha;{\bf R}{\bf R}'}^{\phantom{\dagger}}
      d_{l\sigma{\bf R}}^{\dagger}f_{\alpha{\bf R}'}^{\phantom{\dagger}}+h.c.\Bigr) \nonumber \\
&& +U\sum_{{\bf R}}\sum_{\alpha\beta}f_{\alpha{\bf R}}^{\dagger}f_{\alpha{\bf R}}^{\phantom{\dagger}}
      f_{\beta{\bf R}}^{\dagger}f_{\beta{\bf R}}^{\phantom{\dagger}} \ . \nonumber
\end{eqnarray}
This is a second-quantized Hamiltonian expressed in terms of the creation and annihilation operators of the intrinsic $d$ and $f$ electrons. The spread of the samarium's atomic orbitals into bands is captured by the $d$ and $f$ band dispersions $\xi_{l{\bf k}}$ and $\epsilon_{\alpha{\bf k}}$ respectively, while the hybridization $V_{l\sigma\alpha;{\bf R}{\bf R}'}$ is treated separately. For simplicity, this model neglects the spin-orbit coupling of $d$ electrons and thus regards their spin $\sigma = \lbrace \uparrow,\downarrow \rbrace$ as a good quantum number prior to hybridization. The internal quantum number of $f$ orbitals is the multiplet index $\alpha$ prior to hybridization. Throughout this paper we will work in the units $\hbar=a=1$, where $a$ is the lattice constant and all real-space field operators are dimensionless.

Generally, Kondo materials could have multiple dynamically active $d$ orbitals, here indexed by $l=1,\dots,N_d$, and multiple pairs of $f$ orbitals indexed by $\alpha=1,\dots,N_f$. The band-structure of SmB$_6$ is unfortunately not well known due to fundamental difficulties in dealing with strongly interacting $f$ electrons in ab-intio calculations, as well as the limited energy resolution of experimental approaches such as ARPES. However, at least within this model, a Kondo insulator is possible only if there is no more than one spin-degenerate $d$ orbital ($N_d=1$) and one doublet of $f$ orbitals ($N_f=2$) \footnote{Otherwise, one particular linear combination of $d$ orbitals hybridizes with one similar linear combination of $f$ orbitals, while the other (orthogonal) linear combinations of these orbitals ruin the insulting band-structure by crossing and filling the hybridization gap.}. Therefore, we will drop the index $l$ from now on, and work with a doublet of $f$ orbitals.

The hybridization $V_{l\sigma\alpha;{\bf R}{\bf R}'}^{\phantom{\dagger}}$ mixes $d$ and $f$ orbitals across nearby lattice sites ${\bf R}, {\bf R}'$; nearest neighbor mixing is dominant and yields by itself a simple Fourier transform of the hybridization coupling \cite{Dzero2010, Dzero2012}
\begin{equation}\label{HybCouplings}
V_{\sigma\alpha;{\bf k}}^{\phantom{\dagger}} = 
  \sum_{{\bf R}-{\bf R}'}V_{\sigma\alpha;{\bf R}-{\bf R}'}^{\phantom{\dagger}}\, e^{-i{\bf k}({\bf R}-{\bf R}')} \ ,
\end{equation}
which is characterized by a single energy scale $V_0$:
\begin{equation}\label{HybScale}
V_{\bf k} = \sqrt{\sum_{\alpha\sigma}\left\vert V_{\sigma\alpha{\bf k}}\right\vert ^{2}}
  = V_0 \sqrt{\sin^2(k_x) +\sin^2(k_y) + \sin^2(k_z)}
\end{equation} 

The Coulomb interaction is most effective in the small-bandwidth $f$ band. Occupying a single lattice site with more than one $f$ electron costs potential energy that greatly exceeds their kinetic energy and can be regarded as a high energy state. Instead of formulating a complicated microscopic interaction potential, we capture all important physics by the on-site repulsion $U$ among the $f$ electrons. Furthermore, we approximate $U\to\infty$ by comparison to other energy scales, prohibiting double site occupancy by $f$ electrons in our theory.

\subsection{Slave boson method}\label{secSlaveBoson}

Given that $U$ is very large in (\ref{MinMod}), we cannot directly treat it as a perturbation to the solvable non-interacting part of the Hamiltonian. Instead, we apply the widely used slave boson approximation \cite{PhysRevB.29.3035}. The $f$ electron field operators are first represented as the product of auxiliary slave boson $b$ and slave fermion $\psi$ operators:
\begin{equation}\label{SlaveBoson}
f_{\alpha{\bf R}}^{\dagger}=\psi_{\alpha{\bf R}}^{\dagger}b_{{\bf R}}^{\phantom{\dagger}} \ .
\end{equation}
This introduced redundancy in the degrees of freedom is removed by a local constraint that prohibits double-occupancy of any lattice site by $f$ electrons:
\begin{equation}\label{Constraint}
\sum_{\alpha}\psi_{\alpha{\bf R}}^{\dagger}\psi_{\alpha{\bf R}}^{\phantom{\dagger}}+b_{{\bf R}}^{\dagger}b_{{\bf R}}^{\phantom{\dagger}}
 = 1 \ . 
\end{equation}
If it were possible to handle the constraint exactly, this new representation of dynamics would be completely equivalent to the original one in the $U\to\infty$ limit. Once the double-occupancy is strictly prohibited, we can formally remove the interaction $U$ term from the Hamiltonian. However, uncontrolled but physically motivated approximate implementations of the constraint are necessary in order to make progress. This approach is often used in studies of spin liquids that exhibit spin-charge separation. We associate the charge of an $f$ electron with the fermionic slave fermion, while the slave boson remains neutral.

The simplest level of approximation is the mean-field slave boson theory, which assumes that the slave bosons are condensed and neglects all quantum fluctuations about the condensate. A slave boson condensate is physically justified in SmB$_6$ by the virtue of samarium's mixed valence: electrons in the $f$ orbitals are still mobile, so the slave bosons can be mobile as well and condense at a finite amplitude $0<|B|<1$. The mean-field approximation amounts to replacing the slave boson operators $b_{\bf R}$ by a single complex amplitude $\langle b_{\bf R} \rangle = B$ in (\ref{SlaveBoson}), and also in the local constraint (\ref{Constraint}) provided that we implement it softly (on average):
\begin{equation}\label{ConstraintAvgMF}
\left\langle\sum_{\alpha}\psi_{\alpha{\bf R}}^{\dagger}\psi_{\alpha{\bf R}}^{\phantom{\dagger}}\right\rangle+|B|^2 = 1
\end{equation}
The ensuing mean-field Hamiltonian obtained from (\ref{MinMod}) is formally non-interacting:
\begin{eqnarray}\label{SBMF}
H&&_{\textrm{mf}} = \int\limits_{\textrm{1BZ}}\frac{d^{3}k}{(2\pi)^{3}}\biggl\lbrack
     \sum_{\sigma}\xi_{\bf k}^{\phantom{\dagger}}d_{\sigma{\bf k}}^{\dagger}d_{\sigma{\bf k}}^{\phantom{\dagger}}
   +\sum_{\alpha}\widetilde{\epsilon}_{{\bf k}}^{\phantom{\dagger}}
     \psi_{\alpha{\bf k}}^{\dagger}\psi_{\alpha{\bf k}}^{\phantom{\dagger}} \nonumber \\
&& +\sum_{\alpha\sigma} \Bigl(V_{\sigma\alpha{\bf k}}^{\phantom{\dagger}} B^*
     d_{\sigma{\bf k}}^{\dagger}\psi_{\alpha{\bf k}'}^{\phantom{\dagger}}+h.c.\Bigr)\biggr\rbrack \\
&& +\frac{\mathcal{V}}{a^3}(\epsilon_f'|B|^2-\eta)(1-|B|^2) \nonumber
\end{eqnarray}
($\mathcal{V}$ is the system volume and $a$ is the lattice constant). The slave boson amplitude $|B|$ is determined by minimizing the ground state energy of the mean-field Hamiltonian subject to the soft constraint (\ref{ConstraintAvgMF}). This is typically done self-consistently through a Lagrange multiplier $\eta$, so we added
\begin{equation}
\eta\sum_{{\bf R}}\left(\sum_{\alpha}\psi_{\alpha{\bf R}}^{\dagger}\psi_{\alpha{\bf R}}^{\phantom{\dagger}}+|B|^{2}-1\right)
\end{equation}
to the original Hamiltonian (\ref{MinMod}). The slave-boson condensate renormalizes both the hybridization term and the $f$ electron dispersion via:
\begin{equation}
\widetilde{\epsilon}_{{\bf k}}=(\epsilon_{{\bf k}}-\epsilon_{f}')|B|^{2}+\epsilon_{f}'+\eta \ .
\end{equation}
$\epsilon_f'$ is the momentum-independent part of the intrinsic $f$ electron dispersion. Later on, in Eq.\ref{TightBinding} and numerical calculations of the collective mode spectrum, we will write $\epsilon_f' = \epsilon_f-\mu$, where $\mu$ is the chemical potential (Fermi level) and $\epsilon_f$ is the energy shift of the $f$ orbitals with respect to the $d$ orbitals.

In this work, we do not attempt to self-consistently determine the renormalized spectrum from the microscopic one because the latter is unknown. Instead, we extract the already renormalized low-energy quasiparticle spectrum from the neutron scattering data. The renormalized hybridization bandgap is taken from resistivity measurements, and sets the value of $|B|V_0$ in our model while providing additional constraints on the electron dispersion. Similarly, estimates of the renormalized $f$ electron dispersion determine the energy displacement $\epsilon_f'+\eta$ of the intrinsic $f$ and $d$ bands. Once we pick a band-structure consistent with experimentally available information, we diagonalize (\ref{SBMF}) and compute $\sum_\alpha \langle \psi_\alpha^\dagger \psi_\alpha^{\phantom{\dagger}} \rangle$ in the band-insulating ground state of hybridized electrons. This then directly determines $|B|^2$ through the soft constraint (\ref{ConstraintAvgMF}).

With this procedure in mind, we can readily diagonalize the mean-field Hamiltonian of hybridized two-fold degenerate $d$ and $f$ orbitals:
\begin{equation}
H_{\textrm{mf}} = \sum_{s\lambda}  \int\limits_{\textrm{1BZ}}\frac{d^{3}k}{(2\pi)^{3}} E_{\lambda{\bf k}}^{\phantom{\dagger}}
  \Psi_{s\lambda{\bf k}}^\dagger \Psi_{s\lambda{\bf k}}^{\phantom{\dagger }} \ .
\end{equation}
The mean-field electron spectrum
\begin{equation}\label{SBMFspectrum}
E_{\lambda{\bf k}} = \frac{\xi_{{\bf k}}+\widetilde{\epsilon}_{{\bf k}}}{2}
  +\lambda\sqrt{\left(\frac{\xi_{{\bf k}}-\widetilde{\epsilon}_{{\bf k}}}{2}\right)^{2}+V_{{\bf k}}^{2}|B|^{2}}
\quad,\quad\lambda=\pm 1
\end{equation}
consists of the conduction ($\lambda=1$) and valence ($\lambda=-1$) bands separated by a small bandgap. Both bands are two-fold spin degenerate, ($s=\pm 1$). The hybridized electron operators are:
\begin{equation}
\Psi_{s\lambda{\bf k}}=\frac{(E_{\lambda{\bf k}}-\widetilde{\epsilon}_{{\bf k}})\widetilde{d}_{s{\bf k}}
  +V_{{\bf k}}B\,\widetilde{\psi}_{s{\bf k}}}{\sqrt{(E_{\lambda{\bf k}}-\widetilde{\epsilon}_{{\bf k}})^{2}+V_{{\bf k}}^{2}|B|^{2}}}
\end{equation}
where
\begin{eqnarray}
\widetilde{d}_{s{\bf k}}&=&\frac{d_{\uparrow{\bf k}}+s\,d_{\downarrow{\bf k}}}{\sqrt{2}} \\
\widetilde{\psi}_{s{\bf k}}&=&\frac{1}{V_{\bf k}}\sum_{\alpha}\Bigl(V_{\uparrow\alpha{\bf k}}^{\phantom{*}}
  +s(-1)^{\alpha}V_{\uparrow\bar{\alpha}{\bf k}}^{*}\Bigr)\psi_{\alpha{\bf k}}^{\phantom{\dagger}}\nonumber
\end{eqnarray}
in terms of the hybridization couplings $V_{\sigma\alpha{\bf k}}$ defined in \cite{Dzero2010, Dzero2012}. The symbol $\bar{\alpha}$ labels the orbital $f$ state obtained from the state $\alpha$ upon time-reversal.

It is important to emphasize that the slave boson condensate considered here does not give rise to Goldstone modes (such gapless collective excitations have never been observed in SmB$_6$). There are two reasons for this. First, our slave bosons are electrically neutral, being related to the collective exciton modes that we seek to understand \cite{Nikolic2014b}. The only global U(1) symmetry of the Hamiltonian is the one corresponding to charge conservation. Therefore, a neutral slave boson condensate cannot spontaneously break any symmetry of the Hamiltonian. Its formation can be viewed instead as a result of an explicit ``symmetry'' violation in the Hamiltonian, which allows only gapped collective modes. It should be noted here that the slave boson representation (\ref{SlaveBoson}) of the electron operator introduces an unphysical local (gauge) symmetry in the \emph{slave boson} Hamiltonian, under the transformation $b_{\bf R} \to e^{i\theta_{\bf R}} b_{\bf R}$, $\psi_{\alpha{\bf R}} \to e^{i\theta_{\bf R}} \psi_{\alpha{\bf R}}$. No physical degree of freedom in the original Hamiltonian (\ref{MinMod}) is associated with this fictitious symmetry, so we must ``fix the gauge''. Even though the condensate formally breaks the global part of this symmetry, this cannot correspond to any physical gapless excitations. The second reason for the absence of Goldstone modes lurks in the exact local constraint (\ref{Constraint}). Any propagation of a slave boson current on top of the uniform condensate is correlated with a slave fermion current that propagates in the opposite direction. Fermion currents are gapped in Kondo insulators, so the slave boson currents must be gapped as well. This effect is missed in the averaged constraint (\ref{ConstraintAvgMF}).

\subsection{Neutron scattering and susceptibility}

Inelastic neutron scattering probes magnetic correlations in materials. Since the low-energy scattering processes in Kondo insulators involve electrons with large effective mass, we may assume that neutrons are most sensitive to the electrons' internal degrees of freedom. The neutron scattering cross section is then proportional to the product of the ``transverse'' spin-spin correlation function:
\begin{equation}\label{StructFact}
\mathcal{S}^{\perp}({\bf q},\Omega)=\sum_{{\bf R}}\int\limits_{-\infty}^{\infty}dt\, e^{-i({\bf qR}-\Omega t)}
  \left\langle {\bf S}_{{\bf R}}^{\perp}(t){\bf S}_{0}^{\perp}(0)\right\rangle \ .
\end{equation}
and a momentum-dependent form factor $F({\bf q})$ that encodes the details of scattering from a single ion \cite{SquiresTNS}. The total electron's angular momentum operators ${\bf S}_{{\bf R}}^{\perp}(t)$ at the lattice position ${\bf R}$ and time $t$ (in the Heisenberg picture) are projected on the plane perpendicular to ${\bf q}$. This correlation is also related to the dynamical magnetic susceptibility, which we can obtain by linear response theory. The Zeeman energy of spins in an external magnetic field is
\begin{equation}
V_{\textrm{Z}}(t)=-\sum_{{\bf R}}{\bf B}_{{\bf R}}(t)\boldsymbol{\mu}_{{\bf R}}(t)
  =-\gamma\sum_{{\bf R}}{\bf B}_{{\bf R}}(t){\bf S}_{{\bf R}}(t) \ .
\end{equation}
The response of spins to this perturbation is:
\begin{equation}\label{SpinResp}
\bigl\langle S^i_{{\bf R}}(t)\bigr\rangle=-\gamma\sum_{{\bf R}'}dt'\, \mathcal{D}^{ij}_{{\bf R}-{\bf R}'}(t-t') B^j_{{\bf R}'}(t')
\end{equation}
in terms of the retarded response function tensor
\begin{equation}
\mathcal{D}^{ij}_{{\bf R}-{\bf R}'}(t-t')=-i\bigl\langle\lbrack S^i_{{\bf R}}(t),S^j_{0}(0)\rbrack\bigr\rangle\theta(t) \ ,
\end{equation}
where the superscripts $i,j\in\lbrace x,y,z \rbrace$ label spatial directions. The Fourier transform of (\ref{SpinResp}) thus reveals the proportionality between the dynamical magnetic susceptibility tensor $\chi^{ij}({\bf q},\Omega)$ and $\mathcal{D}^{ij}({\bf q},\Omega)$. We can now express the non-retarded response (\ref{StructFact}) as:
\begin{equation}\label{StructFact2}
\mathcal{S}^{\perp}({\bf q},\Omega) = \left(\delta_{ij} - \frac{q_i q_j}{q^2}\right) \mathcal{S}^{ij}({\bf q},\Omega) \ ,
\end{equation}
where $-i\mathcal{S}^{ij} = \textrm{Re} \lbrace -i\mathcal{S}^{ij}_{\textrm{r}} \rbrace + \textrm{Im} \lbrace -i\mathcal{S}^{ij}_{\textrm{r}} \rbrace \textrm{sign}(\Omega)$ and:
\begin{equation}
\mathcal{S}^{ij}_{\textrm{r}}({\bf q},\Omega) = 
  i\mathcal{D}^{ij}({\bf q},\Omega)=i\sum_{{\bf R}}\int dt\, e^{-i({\bf q}R-\Omega t)}\mathcal{D}^{ij}_{{\bf R}}(t)
\end{equation}
is the retarded ``dynamical structure factor''.

As shown above, the key ingredient of the response theory that relates to the neutron scattering experiment is the spin-spin correlation function. Since the spin operators are obtained from the bilinear products of electron creation and annihilation operators $\Psi^\dagger,\Psi$, we are ultimately interested in the generalized correlation function:
\begin{equation}\label{CorrFunct}
\bigl\langle\Psi_{n{\bf R}}^{\dagger}(t)\Psi_{m{\bf R}}^{\phantom{\dagger}}(t)
  \Psi_{m'0}^{\dagger}(0)\Psi_{n'0}^{\phantom{\dagger}}(0)\bigr\rangle
\end{equation}
The quantum numbers of electrons are here labeled by $n,m$, etc. In the simple case of localized $S=\frac{1}{2}$ spins we would have
\begin{equation}
{\bf S}_{{\bf R}}^{\phantom{\dagger}}(t)=\frac{1}{2}\Psi_{\alpha{\bf R}}^{\dagger}(t)\boldsymbol{\sigma}_{\alpha\beta}^{\phantom{\dagger}}\Psi_{\beta{\bf R}}^{\phantom{\dagger}}(t) \ ,
\end{equation}
where $\boldsymbol{\sigma}$ are Pauli matrices and $\alpha,\beta$ label spin-projection states. However, in our case the electrons of SmB$_6$ are not localized and carry a complicated internal quantum number $n$ shaped by the spin-orbit coupling and crystal fields. This introduces another momentum-dependent factor that contracts the indices of (\ref{CorrFunct}) and converts this simple correlation function to the measured quantity (\ref{StructFact2}). The full form factor has no frequency dependence, so it cannot affect the energy dispersion of any collective mode. However, it modulates the magnitude of the neutron cross section as a function of momenta, especially outside of the first Brillouin zone in relation to the electron wavefunctions at the length scales of a single unit-cell. Note that
\begin{equation}
{\bf S}_{1}^{\perp}{\bf S}_{2}^{\perp}=S_{1}^{x}S_{2}^{x}+S_{1}^{y}S_{2}^{y}=\frac{1}{2}\left(S_{1}^{+}S_{2}^{-}+S_{1}^{-}S_{2}^{+}\right)
\end{equation}
for neutrons whose momentum is ${\bf q}=|{\bf q}|\hat{{\bf z}}$. Thus, the measured quantity $S^{\perp}({\bf q},\Omega)$ reflects spin flips or changes along the direction of ${\bf q}$. Regardless of the magnitude of particle spins, the involved spin changes are always by $\hbar$, which is physically appropriate for neutrons. Thus, neutrons couple to the total internal angular momentum of the material's degrees of freedom in this fashion.

We do not pursue a derivation of the form factor in this work because we are primarily interested in the energy dispersion of the collective mode. Instead, we calculate the generic spin response function:
\begin{eqnarray}\label{Susc}
\chi_{nm,n'm'}(q)&\propto&-i\int\limits_{\textrm{1BZ}}\frac{d^{4}k}{(2\pi)^{4}}\frac{d^{4}k'}{(2\pi)^{4}} \\
  && \times \bigl\langle\Psi_{n,k}^{\dagger}\Psi_{m,k+q}^{\phantom{\dagger}}\Psi_{m',k'+q}^{\dagger}\Psi_{n',k'}^{\phantom{\dagger}}\bigr\rangle
   \nonumber
\end{eqnarray}
whose indices are eventually contracted to yield the dynamical magnetic susceptibility
\begin{equation}\label{Susc2}
\chi(q)=\mathcal{C}_{nm,n'm'}({\bf q})\chi_{nm,n'm'}(q) \ .
\end{equation}
For simplicity, we label the internal electron's mean-field quantum numbers $(s,\lambda)$ by $n,m$, and denote by $k=(\omega,{\bf k}$) and $q=(\Omega,{\bf q})$ the electron's space-time momenta.

Even a non-interacting system contributes to the dynamical magnetic susceptibility through its virtual particle-hole excitations:
\begin{eqnarray}\label{Susc1}
\chi_{0;nm,n'm'}(q) &=& i\int\frac{d^{4}k}{(2\pi)^{4}}G_{0;nn'}(k)G_{0;m'm}(k+q) \nonumber \\
&& -i\rho^{2}\,(2\pi)^{4}\delta(q)\delta_{nm}\delta_{n'm'} \ ,
\end{eqnarray}
The bare hybridized electron propagator in this expression is given by:
\begin{equation}\label{BareElProp}
G_{0;s\lambda,s'\lambda'}(k)=\frac{\delta_{ss'}\delta_{\lambda\lambda'}}{\omega-E_{\lambda{\bf k}}+i\lambda 0^{+}} \ .
\end{equation}
We are interested only in the $q\neq0$ dependence of susceptibilities that reveal the dynamics of collective excitations. The second term of (\ref{Susc1}) involving the local instantaneous (uniform) density $\rho$ of electrons will, therefore, drop out from our analysis. The remaining bare susceptibility yields the well-known Lindhardt function, which we discuss in detail in the following section.

\subsection{Extracting the low-energy quasiparticle spectrum from neutron data}\label{secQSpect}

Before proceeding with a perturbative calculation of the collective mode dispersion in Section \ref{secSB}, we analyze here the Lindhardt function in the mean-field slave boson model. The Lindhardt function describes the incoherent particle-hole excitations; the renormalization of the quasiparticle spectrum by the slave boson condensate will be taken into account, but no collective modes can be captured by this analysis. Nevertheless, a collective mode peak of the magnetic susceptibility will have its largest spectral weight near the regions of the first Brillouin zone where the Lindhardt function is largest (according to the RPA). Therefore, the Lindhardt function provides a useful first comparison relevant to the low-energy quasiparticle band-structure. 

The $T=0$ Lindhardt function (\ref{Susc1}) for the mean-field particle-hole excitations is:
\begin{eqnarray}\label{Lindhardt}
\chi_0(q) &=& i\int\frac{d^{4}k}{(2\pi)^{4}} \frac{1}{\omega-E_{-,{\bf k}}-i0^{+}} \\
&& ~~~~~~~~~~~ \times \frac{1}{\omega+\Omega-E_{+,{\bf k}+{\bf q}}+i0^{+}} \nonumber \\
&=& \int\frac{d^{3}k}{(2\pi)^{3}}\,\frac{1}{\Omega-(E_{+,{\bf k}+{\bf q}}-E_{-,{\bf k}})+i0^{+}} \ , \nonumber
\end{eqnarray}
where the hybridized quasiparticle energy $E_{\lambda,{\bf k}}$ is given by (\ref{SBMFspectrum}) and we ignored the multiplicative factor arising from the contraction of internal degrees of freedom. We use tight-binding dispersions $\xi_{\bf k}$ and $\epsilon_{\bf k}$ for the unhybridized $d$ and $f$ bands respectively:
\begin{eqnarray}\label{TightBinding}
\xi_{{\bf k}}&=&-2t_{d1}\sum_{i}^{x,y,z}\cos(k_{i})
                -2t_{d2}\sum_{i\neq j}^{x,y,z}\cos(k_{i})\cos(k_{j}) \nonumber \\
             && -2t_{d3}\cos(k_{x})\cos(k_{y})\cos(k_{z})-\mu \ , \\
\epsilon_{{\bf k}}&=&-2t_{f1}\sum_{i}^{x,y,z}\cos(k_{i})
                -2t_{f2}\sum_{i\neq j}^{x,y,z}\cos(k_{i})\cos(k_{j}) \nonumber \\
             && -2t_{f3}\cos(k_{x})\cos(k_{y})\cos(k_{z})+\epsilon_f-\mu \ , \nonumber
\end{eqnarray}
up to the third-neighbor hopping. 

The values of $t_{dn}$, $t_{fn}$, $\epsilon_f$ as well as $V_0$ in (\ref{HybScale}) are considered variational parameters. We estimate them from a few known properties of the low-energy quasiparticle band-structure as well as the measured momentum-dependent neutron scattering intensity. 
The magnitude of $d$ and $f$ bandwidths, affecting $t_i$ in (\ref{TightBinding}), and their approximate intersection ($\epsilon_f$) was set by comparison to ARPES measurements \cite{SmB6PastAndPresent}. To maintain an insulating gap under hybridization, $t_{d i}$ and $t_{f i}$ were related by a negative scale factor (Fig.\ref{HybGap}). The value of hybridization was set to induce an indirect band gap of $\approx 19 \textrm{ meV}$ matching the charge gap seen in transport measurements. 

\begin{figure}[t!]
\includegraphics[totalheight=0.25\textheight,viewport= 125 50 725 660, clip]{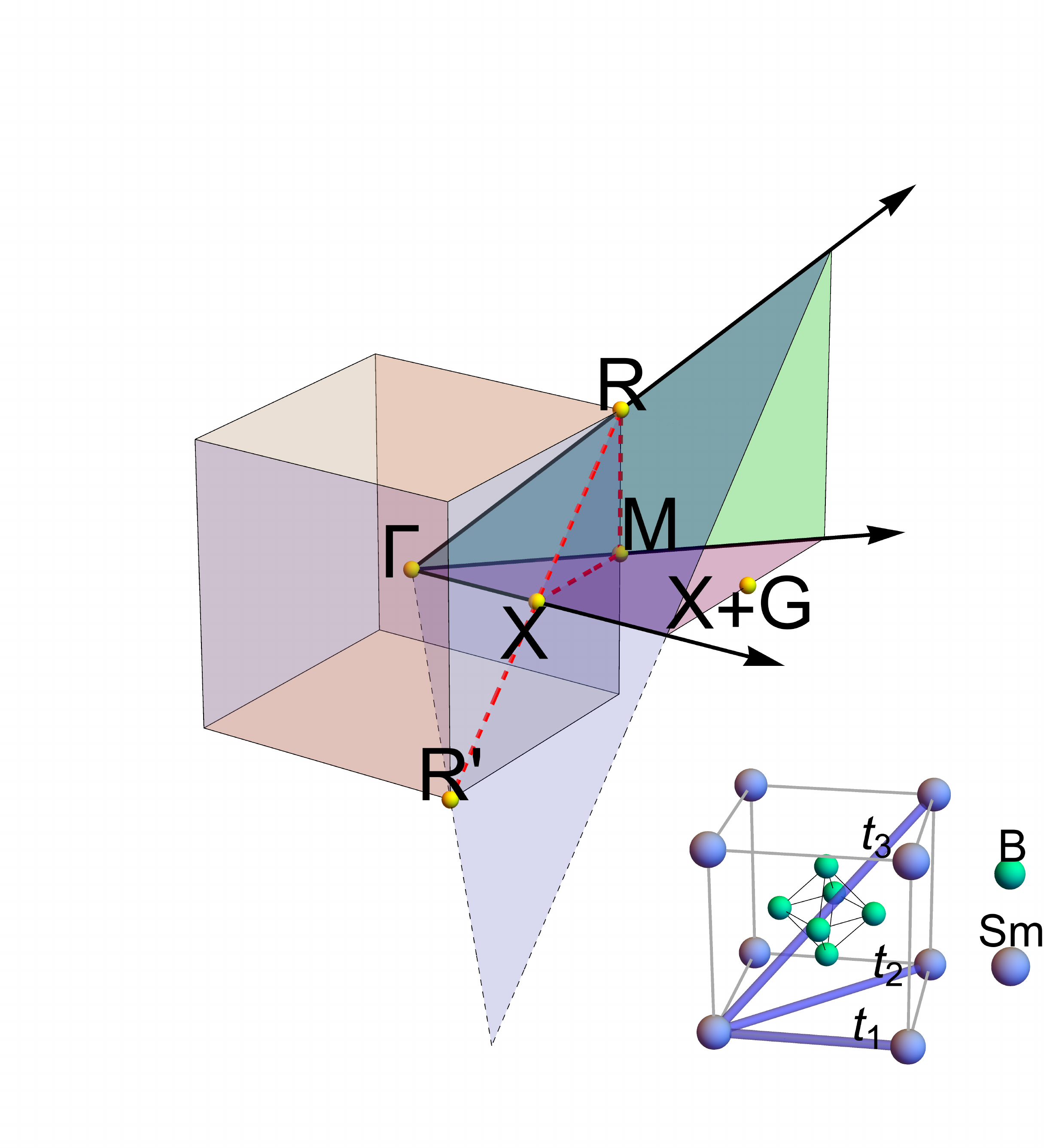}
\caption{ \label{diagrams} Upper left, box denotes the first Brillouin zone, while the wedge is the smallest symmetric portion of momentum space and contains all high symmetry points.  The plane containing $\Gamma$, X and R is duplicated for comparison to Fig.~\ref{qmap}.   Lower right,  SmB$_6$ crystal structure and hopping interactions.  $t_1, t_2, t_3$ are first, second, and third neighbor hopping in the tight-binding model.}
 \end{figure}

In determining the dispersion of the band structure in our minimal model, we considered hopping integrals including first, second, and third-neighbors ($i.e.$ all simple exchanges within the Bravais lattice), using the Lindhardt function (\ref{Lindhardt}) as a guide. Integrating the inelastic neutron spectrum over the range of the dispersion of the observed resonant mode approximates the momentum-space modulation of the low-energy incoherent intensity and is easily calculated via the Lindhardt function.  High energy contributions are not expected to contribute significantly to the bound state and are excluded in this comparison.

The lowest energy particle-hole excitations arise from the quasiparticle dispersion extrema (pointed out in Fig.\ref{HybGap}) that form upon hybridization. Since the Lindhardt function tends to have largest amplitudes at the wavevectors of the lowest energy particle-hole pairs, we can indirectly deduce important features of the quasiparticle spectrum by comparing the Lindhardt function to the neutron scattering intensity. In this manner we can even gain insight on the the band structure topology, irrespective of the model used.  The requisite symmetry of the resonant intensity at momentum transfer X was used to constrain the location of a band inversion to the X (and not M) point of the 1st Brillouin zone in our previous work. Further investigation along this line is in progress, and could potentially be utilized in the study of other Kondo Insulator excitonic phenomena, even in the weakly interacting regime.  The parameter space was extensively sampled, including values previously indicated as relevant for \smb\cite{Legner2014}.  However, dominant third-neighbor hopping showed a close correspondence with INS data, having low-energy resonant intensity peaked near the X and R crystallographic points.  Further supported by the molecular orbital configuration of SmB$_6$ (described above), we chose a third-neighbor dominated band structure.

The guidance provided by the experiments and the Lindhardt function greatly simplified the determination of an optimal quasiparticle spectrum. Our best fit for the 3rd-neighbor-only band-structure parameters is ultimately obtained from the full collective mode calculation and presented in Section \ref{secExp}.

\section{Perturbative calculation of the collective mode spectrum}\label{secSB}

\subsection{Fluctuations beyond the mean-field}

The mean-field slave boson model is incapable of describing any collective modes. In the pursuit of collective modes, we must now consider the dynamics of slave bosons. The full slave boson Hamiltonian is obtained from the Anderson lattice model (\ref{MinMod}) by removing the Coulomb term in favor of the local constraint applied on the slave boson representation (\ref{SlaveBoson}). Small slave boson fluctuations about the condensate can be represented by a dynamical field $\delta b_{\bf R}$:
\begin{equation}\label{SlaveBosonFluct}
b_{{\bf R}}=B+\delta b_{{\bf R}} \ .
\end{equation}
Averaging the local constraint (\ref{Constraint}) then produces:
\begin{equation}\label{ConstraintAvg}
\left\langle\sum_{\alpha}\psi_{\alpha{\bf R}}^{\dagger}\psi_{\alpha{\bf R}}^{\phantom{\dagger}}
  +\delta b_{\bf R}^\dagger\delta b_{\bf R}^{\phantom{\dagger}}\right\rangle +|B|^2 = 1 \ .
\end{equation}
The fluctuations, therefore, shift the condensate amplitude $B$ with respect to the pure mean-field value arising from (\ref{ConstraintAvgMF}), but this shift is small and we neglect it in our calculations. Once the slave boson amplitude is obtained self-consistently or by other means, the sufficiently small fluctuations $\delta b_{\bf R}$ are essentially unconstrained. The ensuing slave boson Hamiltonian expressed in the basis of the mean-field hybridized electrons takes the form:
\begin{widetext}
\begin{eqnarray}\label{SB}
&&H = \sum_{s\lambda}\int\limits_{\textrm{1BZ}}\frac{d^{3}k}{(2\pi)^{3}}\, E_{\lambda{\bf k}}^{\phantom{\dagger}}
  \Psi_{s\lambda{\bf k}}^{\dagger}\Psi_{s\lambda{\bf k}}^{\phantom{\dagger}} +
  \sum_{ss'}\sum_{\lambda\lambda'}\int\limits_{\textrm{1BZ}}\frac{d^{3}k}{(2\pi)^{3}}\frac{d^{3}k'}{(2\pi)^{3}}\left\lbrack
  V_{s\lambda,s'\lambda'}^{\phantom{\dagger}}({\bf k},{\bf k}')
    \Psi_{s\lambda{\bf k}}^{\dagger}\Psi_{s'\lambda'{\bf k}'}^{\phantom{\dagger}}
  \delta b_{{\bf k}'-{\bf k}}^{\dagger}+h.c.\right\rbrack \\
&& ~~ -\sum_{\alpha}\sum_{{\bf R}{\bf R}'}t_{f{\bf R}{\bf R}'}^{\phantom{f}}\left(
     B\delta b_{{\bf R}'}^{\dagger}+B^{\dagger}\delta b_{{\bf R}}^{\phantom{\dagger}}
      +\delta b_{{\bf R}'}^{\dagger}\delta b_{{\bf R}}^{\phantom{\dagger}}\right)
     \psi_{\alpha{\bf R}}^{\dagger}\psi_{\alpha{\bf R}'}^{\phantom{\dagger}}
   +\sum_{{\bf R}}\left\lbrack \epsilon_{f}(1-4|B|^{2})|\delta b_{{\bf R}}^{\phantom{\dagger}}|^{2}
   -\epsilon_{f}(B^{\dagger2}\delta b_{{\bf R}}^{2}+h.c.)+\mathcal{O}(\delta b^{3})\right\rbrack \nonumber
\end{eqnarray}
where
\begin{eqnarray}\label{HybVertex}
&& V_{s\lambda,s'\lambda'}({\bf k},{\bf k}')=a^{\frac{3}{2}}\frac{\lambda B^{*}}{V_{{\bf k}}|B|}\,
    \sqrt{\frac{(E_{\lambda{\bf k}}-\widetilde{\epsilon}_{{\bf k}})^{2}
    +V_{{\bf k}}^{2}|B|^{2}}{\left(\xi_{{\bf k}}-\widetilde{\epsilon}_{{\bf k}}\right)^{2}
  +4V_{{\bf k}}^{2}|B|^{2}}\,\frac{(E_{(-\lambda'){\bf k}'}-\widetilde{\epsilon}_{{\bf k}'})^{2}
    +V_{{\bf k}'}^{2}|B|^{2}}{\left(\xi_{{\bf k}'}-\widetilde{\epsilon}_{{\bf k}'}\right)^{2}
  +4V_{{\bf k}'}^{2}|B|^{2}}} \\ \nonumber
&& ~~~~~~~~~~~~ \times \sum_{\alpha}\lbrack V_{\uparrow\alpha{\bf k}}^{\phantom{*}}V_{\uparrow\alpha{\bf k}'}^{*}
+ ss'V_{\uparrow\alpha{\bf k}}^{*}V_{\uparrow\alpha{\bf k}'}^{\phantom{*}}
+(-1)^{\alpha}\Bigl(s'V_{\uparrow\alpha{\bf k}}^{\phantom{*}}V_{\uparrow\bar{\alpha}{\bf k}'}^{\phantom{*}}
  +sV_{\uparrow\bar{\alpha}{\bf k}}^{*}V_{\uparrow\alpha{\bf k}'}^{*}\Bigr)\rbrack \ .
\end{eqnarray}
\end{widetext}
Note that the factor of $a^{\frac{3}{2}}$, where $a$ is the lattice constant, appears in the definition of $V_{s\lambda,s'\lambda'}({\bf k},{\bf k}')$ because we define the Fourier transforms as:
\begin{equation}
b_{{\bf k}}=a^{\frac{3}{2}}\sum_{{\bf R}}e^{-i{\bf kR}}b_{{\bf R}}
  =a^{-\frac{3}{2}}(2\pi)^{3}\delta({\bf k})B+\delta b_{{\bf k}} \ .
\end{equation}
We also used the relationship
\begin{equation}
V_{\sigma,\alpha;{\bf k}}^{\phantom{\dagger}}=(-1)^{\alpha+(\sigma+1)/2}\, V_{-\sigma,\bar{\alpha};{\bf k}}^{*}
\end{equation}
between the hybridization couplings (\ref{HybCouplings}), which stems from the TR invariance under $\sigma\to-\sigma$, $\alpha\to\bar{\alpha}$, ${\bf k}\to-{\bf k}$. $t_{f{\bf R}{\bf R}'}$ are the tight-binding hopping integrals of the intrinsic $f$ electron taken from (\ref{TightBinding}). Generally, both hybridized electron labels $s$ and $\lambda$ are not good quantum numbers beyond the mean-field approximation.

\begin{figure}
\subfigure[{}]{\includegraphics[height=0.6in]{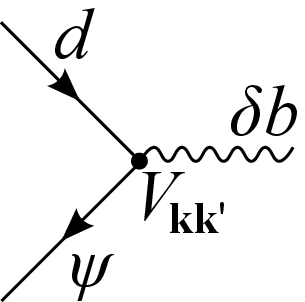}}\hspace{0.4in}
\subfigure[{}]{\includegraphics[height=0.6in]{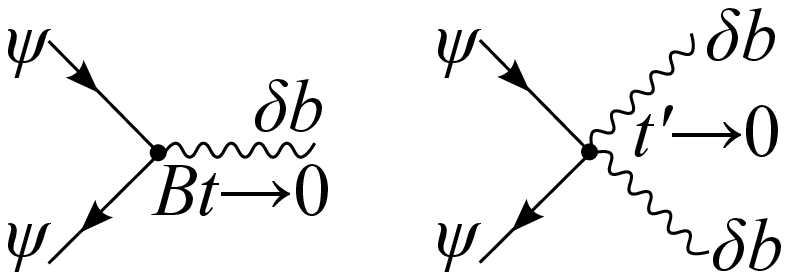}}
\caption{\label{Vertices}The Feynman diagram representations of the interactions between electrons mediated by slave bosons. (a) The dominant hybridization vertex. (b) A leftover from the $f$ electron hopping.}
\end{figure}

The above representation of the Hamiltonian is useful for the perturbative treatment of slave boson fluctuations on top of the condensate. Perturbative processes can be organized using Feynman diagrams; there are three vertices that depict interactions among electrons mediated by slave bosons. The dominant vertex shown in Fig.\ref{Vertices}(a) describes conversion between a slave fermion and a $d$ electron assisted by the emission or absorption of a slave boson. Its coupling constant is determined by the large hybridization energy scale $V_0$. The remaining two vertices shown in Fig.\ref{Vertices}(b) are a leftover from the slave-boson representation of $f$ electron hopping. They are proportional to the intrinsic hopping $t_{f{\bf R}{\bf R}'}$ within the $f$ orbitals. We are safe to neglect these vertices as the narrow bandwidth of $f$ orbitals makes all the hopping constants $t_{f{\bf R}{\bf R}'}$ small.

The bare propagator of hybridized electrons is given by the Green's function (\ref{BareElProp}). The bare slave-boson propagator is made complicated by the formal presence of a condensate. The simplest thing to do is pick an arbitrary condensate phase, say $B\in\mathbb{R}>0$, and separate the fluctuations of $\delta b_{{\bf R}}$ into their longitudinal $\delta b_{\parallel{\bf R}}$ and transverse $\delta b_{\perp{\bf R}}$ parts:
\begin{equation}
\delta b_{{\bf R}}=\delta b_{\parallel{\bf R}}+i\delta b_{\perp{\bf R}}
\quad,\quad\delta b_{\parallel{\bf R}}\,,\,\delta b_{\perp{\bf R}}\in\mathbb{R} \ .
\end{equation}
This can be consistently done in the coherent-state path-integral formulation of this problem, where the complex slave boson field $\delta b_{{\bf R}}$ is easliy decomposed into its real and imaginary parts. We have the following replacements in the Hamiltonian (\ref{SB}):
\begin{eqnarray}
&|&\delta b_{{\bf R}}^{\phantom{\dagger}}|^{2}=\delta b_{\parallel{\bf R}}^{2}+\delta b_{\perp{\bf R}}^{2}
\quad,\quad \\ \nonumber
&B&^{\dagger2}\delta b_{{\bf R}}^{2}+h.c.=2B^{2}\left(\delta b_{\parallel{\bf R}}^{2}-\delta b_{\perp{\bf R}}^{2}\right) \ .
\end{eqnarray}
The longitudinal and transverse slave bosons have different but non-anomalous bare propagators:
\begin{eqnarray}
D'_{0;\parallel}({\bf q},\Omega)&\sim&\frac{2\Omega}{\Omega^{2}-(4\epsilon_{f}B^{2})^{2}+i0^{+}}
\quad,\quad \\ \nonumber
D'_{0;\perp}({\bf q},\Omega)&\sim&\frac{2\Omega}{\Omega^{2}+i0^{+}} \ .
\end{eqnarray}
Both artificially arise from the local slave boson construction (\ref{SlaveBoson}), and formally have flat dispersions obtained from the approximate value of the Lagrange multiplier $\eta \approx \epsilon_f (2B^2-1)$ that leads to (\ref{SB}) without uniform linear terms in $\delta b_{\bf R}$. Longitudinal bosons are clearly gapped and can be ignored (or integrated out) as high energy fluctuations. However, transverse slave bosons appear to be not only flat, but also gapless as a consequence of naively having a condensate. This would be a problem for perturbation theory if it were real. Fortunately, the anomalous slave boson spectrum is a too simplistic outcome of an excessive approximation. A qualitatively more accurate expression for the transverse propagator that takes into account the \emph{hard} local constraint (\ref{Constraint}) is:
\begin{equation}\label{SBtrans}
D_{0;\perp}({\bf q},\Omega) \sim \frac{2\Omega}{\Omega^{2}-\widetilde{E}_{\bf q}^2+i0^{+}} \ ,
\end{equation}
where $\widetilde{E}_{\bf q}$ is the energy of some representative particle-hole excitation that carries the momentum $\bf q$. As we emphasized earlier, the slave bosons cannot have Goldstone modes. There is a small gap to creating collective modes of slave bosons because they must pull backward the slave fermions on their path. This gap is of the order of the electron bandgap in SmB$_6$, implying that the full slave boson propagator is dominated by its self-energy, as we will show in the next section. Therefore, we will be able to neglect the bare slave boson propagator whose mathematical form is hard to express. The slave boson dynamics is entirely ``effective''.

Upon the separation of longitudinal and transverse slave bosons, the simplified Hamiltonian (\ref{SB}) can be written as:
\begin{eqnarray}\label{PertHamiltonian}
H &=& \int\limits_{\textrm{1BZ}}\frac{d^{3}k}{(2\pi)^{3}} \left\lbrack
      \sum_{n}E_{n{\bf k}}^{\phantom{\dagger}}\Psi_{n{\bf k}}^{\dagger}\Psi_{n{\bf k}}^{\phantom{\dagger}}
  + \widetilde{E}_{{\bf q}}^{\phantom{\dagger}} (\delta b_{{\bf q}}^{\perp})^2 \right\rbrack \\
&& + \sum_{nn'}\int\limits_{\textrm{1BZ}}\frac{d^{3}k}{(2\pi)^{3}}\frac{d^{3}k'}{(2\pi)^{3}} V_{nn'}^{\perp}({\bf k},{\bf k}') 
    \Psi_{n{\bf k}}^{\dagger}\Psi_{n'{\bf k}'}^{\phantom{\dagger}} \delta b_{{\bf k}'-{\bf k}}^{\perp} \ , \nonumber
\end{eqnarray}
where
\begin{equation}
V_{nn'}^{\perp}({\bf k},{\bf k}') = -iV_{nn'}^{\phantom{*}}({\bf k},{\bf k}') +iV_{n'n}^{*}({\bf k}',{\bf k}) \ .
\end{equation}
From now on, we will label the mean-field quantum numbers of hybridized electrons by a single symbol, $n=(s,\lambda)$. This is the Hamiltonian that we will work with. It neglects all vertices that describe pure slave fermion scattering, as well as the longitudinal slave boson fluctuations. Note that $V_{nn'}^{\perp}({\bf k},{\bf k}')=V_{n'n}^{\perp *}({\bf k}',{\bf k})$ is symmetric under the exchange of electron labels even though $V_{nn'}^{\phantom{*}}({\bf k},{\bf k}')$ is not.

\subsection{Renormalized slave boson propagators and ladder diagram vertex corrections}

\begin{figure}
\includegraphics[height=0.6in]{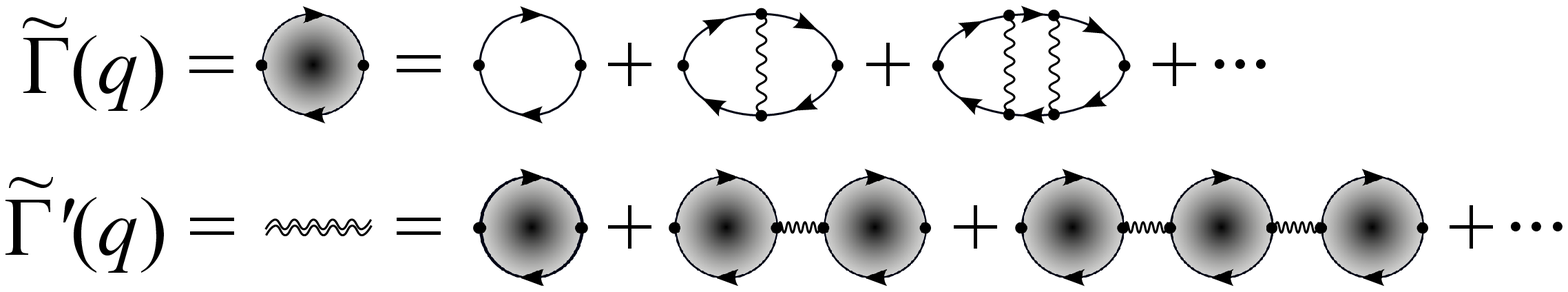}
\caption{\label{LadderChain}The ladder $\widetilde{\Gamma}(q)$ and chain $\widetilde{\Gamma}'(q)$ diagrams that lead to the renormalized susceptibility $\chi(q)$. The wiggly lines represent the slave boson propagators $D(q)$, while the fermion bubbles represent $\Pi(q)$.}
\end{figure}

In order to perturbatively calculate the spin response function (\ref{Susc}) we must consider interactions among electrons mediated by the slave boson fluctuations. Slave bosons are exchanged between electrons in various processes; the character of such interactions is determined by the full transverse slave boson propagator $D(q)$. The particle-hole bubble diagram $\Pi(q)$ provides a self-energy correction to this propagator:
\begin{eqnarray}
&& D(q)=\frac{1}{D_{0;\perp}^{-1}(q)+\Pi(q)} \approx \frac{2\Omega}{\Omega^{2}-\widetilde{E}_{{\bf q}}^{2}+2\Omega\Pi(q)+i0^{+}} \nonumber \\
&& ~~~~~~ \xrightarrow{\Omega=\widetilde{E}_{{\bf q}}+\delta\Omega} \frac{1}{\Pi({\bf q})}
   -\frac{\delta\Omega}{\Pi^{2}({\bf q})}
   +\frac{\Pi+2\widetilde{E}_{{\bf q}}}{2\widetilde{E}_{{\bf q}}^{\phantom{2}}\Pi^{3}({\bf q})}\delta\Omega^{2}+\cdots \nonumber \\
&& ~~~~~~ \xrightarrow{\delta\Omega\ll\Pi({\bf q})}\Pi^{-1}({\bf q}) \ . 
\end{eqnarray}
We used the estimate (\ref{SBtrans}) of the bare transverse slave boson propagator, and applied a few approximations that amount to neglecting this bare propagator in the final result. These approximations are justified in the sense that the full slave boson propagator provides the backbone for the collective mode of SmB$_6$ seen by neutron scattering (see Fig.\ref{LadderChain}). The mode lives at the energy of about $14 \textrm{ meV}$, which is about $|\delta\Omega| \sim 2-5 \textrm{ meV}$ below the quasiparticle bandgap $\widetilde{E}_{{\bf q}} \approx 19 \textrm{ meV}$. We will discover from the comparison with the neutron data that the energy scale of the self-energy correction $\Pi({\bf q}) \sim 1 \textrm{ eV}$ is indeed much larger than $|\delta\Omega|$ as assumed above. 

\begin{figure}
\subfigure[{}]{\includegraphics[height=0.9in]{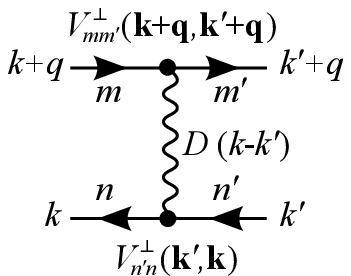}}
\subfigure[{}]{\includegraphics[height=0.9in]{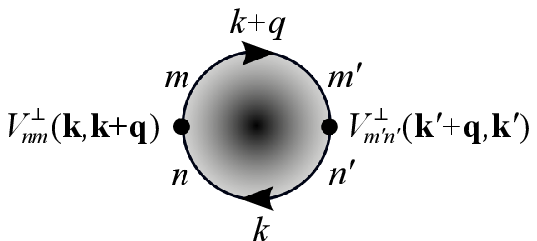}}
\caption{\label{FeynmanNotation}The labels of vertices and propagators in bubble diagrams and their vertex corrections.}
\end{figure}

The bare bubble diagram is calculated as:
\begin{widetext}
\begin{eqnarray}\label{BareBubble}
\Pi_0(q)&=&i\int\frac{d^{4}k}{(2\pi)^{4}}V_{nm}^{\perp}(k,k+q)V_{m'n'}^{\perp}(k+q,k) 
     G_{0;mm'}(k+q)G_{0;n'n}(k) \\
&=& -\int\frac{d^{3}k}{(2\pi)^{3}}\,\frac{\theta(-E_{n,{\bf k}})-\theta(-E_{m,{\bf k}+{\bf q}})}
     {\Omega+E_{n,{\bf k}}-E_{m,{\bf k}+{\bf q}}+i0^{+}\textrm{sign}(E_{m,{\bf k}+{\bf q}})-i0^{+}\textrm{sign}(E_{n,{\bf k}})}
     V_{nm}^{\perp}({\bf k},{\bf k}+{\bf q})V_{mn}^{\perp}({\bf k}+{\bf q},{\bf k}) \nonumber \ .
\end{eqnarray}
\end{widetext}
However, no exciton pairing is attained until a set of vertex corrections to the bubble is taken into account. The needed vertex corrections capture the exchange of an arbitrary number of slave bosons between the particle and the hole, whose propagation is represented by the opposite fermion lines of the bubble diagram. The sum of all such ladder diagrams, visualized in Fig.\ref{LadderChain}, can be calculated from the Dyson equation for a two-body correlation function $\Gamma_{nm{\bf k},n'm'{\bf k}'}(q)$:
\begin{equation}\label{Dyson}
\Gamma=\Gamma_{0}^{\phantom{\dagger}}+\Gamma_{0}^{\phantom{\dagger}}U\Gamma=\Bigl(\Gamma_{0}^{-1}-U\Bigr)^{-1} \ .
\end{equation}
This equation represents $\Gamma_{nm{\bf k},n'm'{\bf k}'}$ as a matrix $\Gamma$ whose rows are indexed by $nm{\bf k}$ and columns by $n'm'{\bf k}'$. The bare two-body correlator is:
\begin{equation}\label{BareGamma}
\Gamma_{0;nm,n'm'}(k,k')=iG_{mm'}(k+q)G_{n'n}(k)\,(2\pi)^{4}\delta(k-k') \ ,
\end{equation}
while its self-energy correction, shown in Fig.\ref{FeynmanNotation}(a), is:
\begin{equation}\label{GammaSelfEnergy}
U_{nm,n'm'}(k,k')=V_{mm'}^{\perp}(k+q,k'+q)V_{n'n}^{\perp}(k',k)D(k-k') \ .
\end{equation}
Matrix multiplication in (\ref{Dyson}) automatically performs momentum integrations inside loops of the bubbles with vertex corrections. The renormalized bubble diagram, depicted in Fig.\ref{FeynmanNotation}(b), must also contract the external electron propagators of the two-body correlator:
\begin{eqnarray}
\Pi(q)&=&\int\frac{d^{4}k}{(2\pi)^{4}}\frac{d^{4}k'}{(2\pi)^{4}} \Gamma_{nm{\bf k},n'm'{\bf k}'}(q) \\ \nonumber
  && ~~~~~~~~~~~~~~~ \times V_{nm}^{\perp}(k,k+q) V_{m'n'}^{\perp}(k'+q,k') \ .
\end{eqnarray}
This is compatible with (\ref{BareBubble}) if the full two-body correlation is replaced by the bare one (\ref{BareGamma}).

Note that (\ref{GammaSelfEnergy}) contains the renormalized slave boson propagator $D(q)$, which in turn has to be calculated from a Dyson equation that ultimately depends on (\ref{GammaSelfEnergy}). Ideally, this problem should be solved self-consistently. Furthermore, electron propagators have their own self-energy corrections that depend on the renormalized $D(q)$. The full self-consistent treatment is prohibitively complicated. However, our previous INS experiment provides insight into the renormalized dynamics. We need not carry out the full self-consistent calculation since we can glean the values of various renormalized quantities from experiments. For example, we don't calculate the electron self-energy corrections since the spectrum we indirectly extract from the experiment is already renormalized.

\subsection{Practical approximations}\label{secPract}

The Dyson equation (\ref{Dyson}) for the two-body correlator $\Gamma$ is extremely complicated because its self-energy (\ref{GammaSelfEnergy}) has a non-trivial dependence on the momentum transfer $k-k'$. Equations of this kind can be solved analytically only in the absence of such a momentum transfer dependence, as in the RPA calculation of Hubbard model instabilities \cite{WenQFT2004}. Otherwise, this problem is as difficult as an equivalent non-local Schrodinger equation in the four-dimensional space-time, which in principle could be constructed from the Bethe-Salpeter equation if we knew the analytic form of the renormalized slave boson propagator. Given that even the non-interacting part of our Hamiltonian is fairly complex, the problem at hand could be approached only numerically, but this is computationally demanding in the four-dimensional space-time. Therefore, we are forced to make further approximations to make progress.

Of primary concern is the reduction of (\ref{GammaSelfEnergy}) to an expression that does not depend on momentum transfer. We must ignore momentum transfers without losing the important aspects of the momentum dependence in $U_{nm,n'm'}(k,k')$. A closer inspection shows that the momentum dependence of $U_{nm,n'm'}(k,k')$ filters out certain regions of the first Brillouin zone depending on the combination of internal indices $n,m,\dots$ as a consequence of the fact that the vertex microscopically converts an $f$ electron to a $d$ electron. We describe this restriction using the mean-field eigenstate labels $n$ of the hybridized electrons. The filtering effect is embedded in the electron's hybridization vertex, and can be qualitatively extracted by a factorization:
\begin{equation}\label{HybVertex2}
V_{nn'}^{\perp}({\bf k},{\bf k}')=X_{n}({\bf k})X_{n'}({\bf k}')\bar{U}_{nn'}({\bf k},{\bf k}') \ ,
\end{equation}
where $X_{n}({\bf k})$ and $\bar{U}_{nn'}({\bf k},{\bf k}')$ are functions to be determined. It is in our interest to construct a function $\bar{U}_{nn'}({\bf k},{\bf k}')$ with featureless momentum dependence across the entire first Brillouin zone. Then, we may be able to entirely neglect its momentum dependence and replace it by a set of representative numbers $\bar{U}_{nn'}$. Motivated by the expected flat dispersion of the collective mode (which we indeed find both experimentally and theoretically), we may also neglect the momentum dependence of the slave boson propagator $D(q)$ in (\ref{GammaSelfEnergy}). Unfortunately, we must ignore its frequency dependence as well, despite having no true justification: $D(q)$ should have a pole in the frequency range that we scrutinize, given that we expect a low-energy collective mode. Such a pole is at least unlikely to have a strong effect on the frequency and momentum dependence of $\Gamma(q)$ that we are primarily interested in. With these approximations $\bar{U}_{nn'}({\bf k},{\bf k}') \to \bar{U}_{nn'}$ and $D(q) \to D$, we can factorize the self-energy part (\ref{GammaSelfEnergy}) of the two-body correlator:
\begin{eqnarray}
U_{nm,n'm'}(k,k')\approx && X_{m}({\bf k}+{\bf q})X_{m'}({\bf k}'+{\bf q}')\times \\ \nonumber
&&\bar{U}_{mm'} X_{n'}({\bf k}')X_{n}({\bf k})\bar{U}_{n'n} \, D \ .
\end{eqnarray}
Let us now redefine (\ref{BareGamma}) and (\ref{GammaSelfEnergy}) so that the factors $X_{n}({\bf k})$ are moved from the self-energy to the modulated propagator:
\begin{eqnarray}\label{ModProp}
\widetilde{G}_{nn'}^{\phantom{l}}(k)&=&X_{n}(k)G_{nn'}(k)X_{n'}(k) \\
\widetilde{U}_{nm,n'm'}&=&\bar{U}_{mm'}\bar{U}_{n'n}D \nonumber
\end{eqnarray}
Since this removes the momentum dependence from the interaction mediators, the internal momenta on the loops of the ladder diagrams can be independently integrated out:
\begin{equation}\label{ModSusc}
\widetilde{\Gamma}_{0;nm,n'm'}(q)=i\int\frac{d^{4}k}{(2\pi)^{4}}\widetilde{G}_{mm'}(k+q)\widetilde{G}_{n'n}(k) \ ,
\end{equation}
and the Dyson equation (\ref{Dyson}) is dramatically simplified:
\begin{equation}\label{SimpleDyson}
\widetilde{\Gamma}=\widetilde{\Gamma}_{0}^{\phantom{\dagger}}+\widetilde{\Gamma}_{0}^{\phantom{\dagger}}\widetilde{U}\widetilde{\Gamma}
  =\Bigl(\widetilde{\Gamma}_{0}^{-1}-\widetilde{U}\Bigr)^{-1} \ .
\end{equation}
The quantities $\widetilde{\Gamma}$, $\widetilde{\Gamma}_0$ and $\widetilde{U}$ are finite-dimensional matrices indexed by the discrete indices $(nm)$, where $n$ and $m$ take four possible values each. The matrices $\widetilde{\Gamma}$ and $\widetilde{\Gamma}_0$ depend on the 4-momentum $q$.

An inspection of (\ref{HybVertex}) suggests a choice:
\begin{equation}\label{Xfunct}
X_{n}({\bf k})=\sqrt{\frac{(E_{\lambda{\bf k}}-\widetilde{\epsilon}_{{\bf k}})^{2}+V_{{\bf k}}^{2}|B|^{2}}
  {\left(\xi_{{\bf k}}-\widetilde{\epsilon}_{{\bf k}}\right)^{2}+4V_{{\bf k}}^{2}|B|^{2}}}
\end{equation}
because these functions have strong momentum dependence. The corresponding function of momenta:
\begin{equation}
\bar{U}_{nn'}({\bf k},{\bf k}') = \frac{V_{nn'}^{\perp}({\bf k},{\bf k}')}{X_{n}({\bf k})X_{n'}({\bf k}')}
  \xrightarrow{\textrm{approx.}}{}\bar{U}_{nn'} = \textrm{const.}
\end{equation}
is acceptably featureless across the first Brillouin zone: it has some rapid variations, but does not filter out or bias any significant region of the Brillouin zone. Given that it has some momentum dependence, it is not clear how to find the best substitution by the phenomenological constants $\bar{U}_{nn'}$. These constants have the units of $a^{\frac{3}{2}} \times$energy and their order of magnitude is set by the hybridization energy scale $V_0$. We will indirectly fit the values of $\bar{U}_{nn'}$ in our numerical calculations by trying to match the theoretical collective mode dispersion to the experimentally observed one. In this sense, we will read out the renormalized values of various parameters from the experiment instead of trying to calculate them from the unknown microscopic values. At the same time we will fit the representative value $D$ of the slave boson propagator. This freedom of fitting mitigates possible issues behind neglecting the frequency dependence of $D(q)$.

\subsection{The calculation of collective mode dispersion and spectral weight}

We are now ready to calculate the dynamical susceptibility (\ref{Susc}) whose appropriate contraction (\ref{Susc2}) is relevant to the neutron scattering experiment.  The precise form factors $\mathcal{S}_{nm,n'm'}({\bf q})$ involved in the contraction are irrelevant for the mode dispersion, and we pick an arbitrary contraction that simplifies our calculation. A convenient choice is naively:
\begin{equation}
\chi(q)\to\textrm{tr}\left\lbrack\widetilde{\Gamma}(q)\right\rbrack \ ,
\end{equation}
which takes the contributions of all collective modes evenly. The measured mode dispersions correspond to the poles of $\widetilde{\Gamma}(q)$.
However, this incomplete approximation predicts too broad dispersion of the collective mode; we rectify this by taking into account the fluctuations in which a particle and a hole are repeatedly created and recombined as the exciton propagates. The ladder diagrams of the contracted two-body correlator $\widetilde{\Gamma}$ include only the interaction between a propagating electron and a hole, so now all chains of such ladders shown in Fig.\ref{LadderChain} must be summed up. This fixes the bandwidth of the collective mode. We therefore extend the calculation:
\begin{equation}\label{Susc3}
\chi(q)\to\textrm{tr}\left\lbrack\widetilde{\Gamma}'(q)\right\rbrack \ ,
\end{equation}
where
\begin{equation}\label{FixedCorrel}
\widetilde{\Gamma}'(q)=\widetilde{\Gamma}(q)+\widetilde{\Gamma}(q)\widetilde{U}'(q)\widetilde{\Gamma}(q)
\end{equation}
and
\begin{eqnarray}
\widetilde{U}'(q)&=&|\bar{U}\rangle D(q)\langle\bar{U}| 
  \quad,\quad \\ \nonumber 
D^{-1}(q)&=&\Pi(q)=\langle\bar{U}|\widetilde{\Gamma}'(q)|\bar{U}\rangle  \quad,\quad \\ \nonumber 
\langle nm|\bar{U}\rangle&=&\bar{U}_{nm}^{*} \ .
\end{eqnarray}
Here we defined the vector $|\bar{U}\rangle$ from the phenomenological parameters $\bar{U}_{nm}$ in order to set up a convenient matrix representation. Note that the matrix $\widetilde{U}'(q)$ already contains the infinite sum of ladder diagrams, which comes from summing up the self-energy corrections $\Pi(q)$ to the slave-boson propagator $D(q)$. This is why only two terms formally appear in (\ref{FixedCorrel}), instead of infinitely many as in the expanded Dyson equation. Also note that the momentum and frequency dependence of $D(q)$ is taken into account here, in contrast to the approximate ladder diagram calculation (\ref{SimpleDyson}) where we had to neglect it. This turns out to be essential for the fairly flat dispersion of the collective mode. We would ideally want to identify $\widetilde{\Gamma}'(q) \to \widetilde{U}'(q)$ since they correspond to the same sum of diagrams. However, this is not possible with the present level of approximations that tap into the momentum dependence of the slave-boson propagators; (\ref{FixedCorrel}) is a practical compromise.

Now, the measured mode dispersions correspond to the poles of $\widetilde{\Gamma}'(q)$. We can write:
\begin{eqnarray}\label{ModeShape}
\widetilde{\Gamma}'(q)&\approx&\left\lbrack \frac{1}{\Omega-\mathcal{E}_{{\bf q}}+i0^{+}}
    -\frac{1}{\Omega+\mathcal{E}_{{\bf q}}-i0^{+}} \right\rbrack  
      A_{{\bf q}}|\xi_{{\bf q}}\rangle\langle\xi_{{\bf q}}| \nonumber \\ 
&=& \frac{2\mathcal{E}_{{\bf q}}A_{{\bf q}}}{\Omega^{2}-\mathcal{E}_{{\bf q}}^{2}+i0^+}|\xi_{{\bf q}}\rangle\langle\xi_{{\bf q}}|
\end{eqnarray}
in the vicinity of a pole, where $\mathcal{E}_{{\bf q}}$, $A_{\bf q}$ and $|\xi_{{\bf q}}\rangle$ are the mode's energy, spectral weight and eigenvector respectively. The poles of $\widetilde{\Gamma}'(q)$ are zeroes in the frequency dependence of:
\begin{equation}
\widetilde{\Gamma}'^{-1}(q)=\left\lbrack 1+\widetilde{U}'(q)\frac{1}{\widetilde{\Gamma}_{0}^{-1}(q)-\widetilde{U}}\right\rbrack^{-1}
  \!\!(\widetilde{\Gamma}_{0}^{-1}(q)-\widetilde{U}) \ ,
\end{equation}
which follows from (\ref{SimpleDyson}) and (\ref{FixedCorrel}). We look for the zeroes by solving the eigenproblem
\begin{equation}
\widetilde{\Gamma}'^{-1}(\Omega,{\bf q})|\xi_{{\bf q}}\rangle=g|\xi_{{\bf q}}\rangle
\end{equation}
and varying $\Omega$ until we find the eigenvalue
\begin{equation}
g \approx \frac{\Omega^{2}-\mathcal{E}_{{\bf q}}^{2}}{2\mathcal{E}_{{\bf q}}A_{{\bf q}}} \to 0 \ .
\end{equation}
This procedure immediately reveals the collective mode energy $\Omega \to \mathcal{E}_{\bf q}$. At the same time, we obtain the mode's spectral weight:
\begin{equation}\label{SpecWeight}
A_{{\bf q}}=\lim_{\Omega\to\mathcal{E}_{{\bf q}}}\left(\frac{\Omega^{2}-\mathcal{E}_{{\bf q}}^2}{2\mathcal{E}_{{\bf q}}}
  \langle\xi_{{\bf q}}|\widetilde{\Gamma}'({\bf q},\Omega)|\xi_{{\bf q}}\rangle\right) \ .
\end{equation}
The spectral weight determines the imaginary part of the frequency integral of $\chi(q)$. This is related to the integrated intensity of the scattered neutron beam in the experiment. The relationship is not a perfect proportionality because of the momentum-dependent spin form factors that we do not calculate in this paper.

The above derivations that led to the mode dispersion $\mathcal{E}_{\bf q}$ are much more complicated than the actual numerical calculation. The main microscopic ingredient of the calculation is $\widetilde{\Gamma}_0(q)$ obtained from (\ref{ModSusc}). This is nothing but the Lindhardt function with modified electron propagators (\ref{ModProp}) instead of the bare propagators. It captures the dynamics of unbound particle-hole pairs in the spectrum that is renormalized both by the condensate and fluctuations of the slave bosons. The details of the realistic (renormalized) band structure enter $\widetilde{\Gamma}_0(q)$ through the hybridized electron energies $E_{n{\bf q}}$. On the other hand, the matrices $\widetilde{U}$ and $\widetilde{U}'$ are obtained from the fitted phenomenological parameters $\bar{U}_{nm}$ and $D$. They have the units of energy and an overall magnitude comparable to the hybridization energy scale $V_0$. For simplicity, we restrict ourselves to a subset $U_{s\lambda,s'\lambda'} = \delta_{ss'} \lbrack U_1\delta_{\lambda\lambda'} + U_2 (1-\delta_{\lambda\lambda'}) \rbrack$ spanned by two fitting parameters $U_1, U_2$, in addition to $D$.

\section{Comparison to Experiment}\label{secExp}
\begin{figure}
\includegraphics[totalheight=0.3\textheight,viewport= 45 30 330 365, clip]{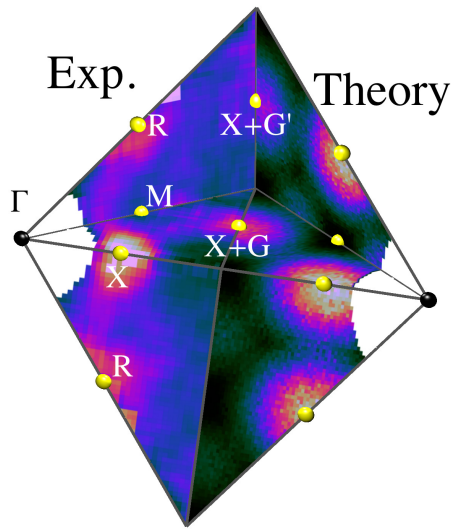}
\caption{ \label{qmap2} Energy integrated intensity of high symmetry planes diagramed in fig.~\ref{diagrams}.  Left, experimental result integrated from $12-16 \textrm{ meV}$.  Right, mirrored calculation from the Lindhardt function (\ref{Lindhardt}) integrated from $10-20 \textrm{ meV}$.  The bare spectrum is inherently incoherent and requires a greater integration range.  The intensity of the calculation is modulated by the 5$d$ form factor to phenomenologically account for the unknown microscopic form factor.}
\end{figure}

The magnetic feature observed in low-energy neutron experiments on SmB$_6$ \cite{Fuhrman2014} is a resolution-limited peak at energies $\approx 13-15 \textrm{ meV}$, below the charge-gap ($\approx 19 \textrm{ meV}$) with dispersion of $<2 \textrm{ meV}$.  The dispersion has local energy minima at high-symmetry X and R points of the first Brillouin zone, $(\frac{1}{2}, 0, 0)$ and $(\frac{1}{2},\frac{1}{2},\frac{1}{2})$ RLU, respectively. The neutron scattering intensity is largely confined to these regions of the first Brillouin zone. This provides many restrictions of the model parameters in our theoretical calculation of the mode spectrum.

In the non-interacting hybridized model, the spectrum of the spin-exciton is incoherent and described by the Lindhardt function.  However, the energy profile of the slave-boson renormalized bound state we calculate in the vicinity of a pole ($i.e.$ where the peak is observed) is given by (\ref{ModeShape}).  This is an infinitely sharp Lorentzian for which any experimental observation thereof would appear as a delta function convoluted with the instrumental resolution function (small thermal broadening can also occur due to interactions among the thermally-excited mode quanta).  This matches the experimental result for all momenta.  This resolution-limited effect indicates the absence of intrinsic damping, a consequence of the mode's energy being below the charge gap where no quasiparticle decay channels are present.  Transport experiments reveal the charge gap to be $\approx 19  \textrm{ meV}$, while the collective mode is observed in neutron experiments at $\approx 14 \textrm{ meV}$ \cite{Fuhrman2014}.  

The mode seen in experiments has a small dispersion of $<2 \textrm{ meV}$.  The calculated mode dispersion $\mathcal{E}_{\bf q}$ can be directly compared with the experiment, since it inherits its energy scale from the experimentally constrained spectrum $E_{n{\bf q}}$ of hybridized electrons.  The dispersion in our calculation is order of magnitude correct ($\approx 6 \textrm{ meV}$).  However, it is unlikely that the minimal model we have based our calculations on perfectly captures the fully renormalized band structure; an additional refinement of the phenomenological parameter values may increase this agreement.

The most-scattered neutrons on SmB$_6$ are those that transfer momentum of the X and R points in the 1st Brillouin zone. This corresponds to peaks of the dynamical susceptibility at these wavevectors, and typically reveals the wavevectors where the particle-hole excitations have the lowest energy. These wavevectors are qualitatively visible even in the particle-hole spectrum of an equivalent non-interacting system with the same band-structure, shown in Fig.~\ref{qmap2}.  As we pointed out in Section \ref{secQSpect}, a simple tight-binding model with dominant third neighbor hopping has the lowest energy particle-hole pairs that congregate precisely to the X and R points of the 1st Brillouin zone.  For momentum transfer near these points there is a larger energy gap, introducing a dispersion upwards in energy, as seen in the experiment.  The hopping in the $d$ band is estimated from the experimentally known bandwidth, while the hopping in the $f$ band must have a much smaller value and opposite sign in order to produce a bandgap (i.e. an insulating behavior). The hybridization energy scale $V_0$ is largely determined by the size of the bandgap (charge gap), which is known from transport measurements.

\begin{figure}\label{qeslices}
\includegraphics[totalheight=.19\textheight]{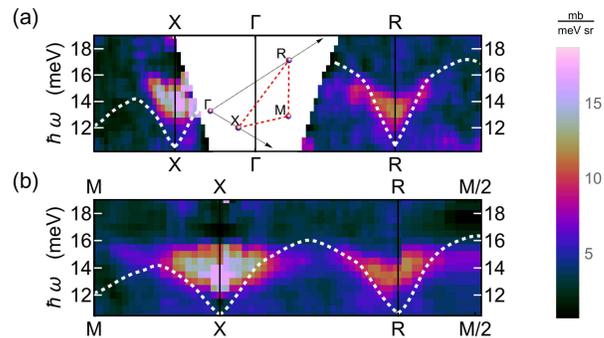}
\caption{ \label{qmap} Energy-resolved neutron scattering intensity along high-symmetry directions of the Brillouin zone.  Dashed line is the perturbative slave-boson dispersion.}
 \end{figure}

We obtained concrete values of all model parameters by trying to match the calculated and measured collective mode dispersions in the entire first Brillouin zone (where ever neutron data is available). Fig.\ref{qmap} compares the theoretical and experimental dispersions for the best simple set of parameters we found so far: $t_{f1} = t_{f2} = t_{d1} = t_{d2} = 0$, $t_{f3} = 0.171 \textrm{ eV}$, $t_{d3} = -0.132 \textrm{ eV}$, $\epsilon_f = 0.035 \textrm{ eV}$, $V_0 = 0.915 \textrm{ eV}$ in the model given by (\ref{MinMod}), (\ref{HybScale}) and (\ref{TightBinding}). These are the ``microscopic'' values. A small amplitude $|B| = 0.077$ of the slave boson condensate obtained from (\ref{ConstraintAvgMF}) significantly renormalizes the dispersion of quasiparticles. The observable bandwidth of $f$ electrons is set by $|B|^2 t_{f3} \sim 1.0 \textrm{ meV}$, and the $19 \textrm{ meV}$ hybridization bandgap is controlled by the energy scale $|B|V_0 \sim 70.4 \textrm{ meV}$ of the same order of magnitude \footnote{The slave boson method sees the bandgap as being proportional to the slave boson amplitude $|B|$. If the temperature is raised, $|B|$ decreases and the bandgap shrinks in agreement with experimental observations.}. The renormalized quasiparticle band-structure corresponds to that of a strong topological insulator \cite{Fuhrman2014}.

The calculation also involves a fit of the three phenomenological parameters that we described at the end of the previous section: $U_1 = 1.41\; a^{\frac{3}{2}}\; \textrm{ eV}$, $U_2 = 1.81\; a^{\frac{3}{2}}\; \textrm{ eV}$ and $D = 5.54 \textrm{ eV}^{-1}$ ($a$ is the lattice constant). Note that the independently obtained fits of $U_1$ and $U_2$ in the units $a=1$ are both comparable with the scale of $V_0$, which is just as expected from the perturbation theory point of view (see Section \ref{secPract}). This supports the confidence that the slave boson perturbation theory and this fitting procedure can capture the correct physics despite being an uncontrolled approximation.

Our calculation of the spectral weight $A_{\bf q}$ is related to the scattered neutron intensity, but cannot be directly compared with it. $A_{\bf q}$ in (\ref{SpecWeight}) is contaminated by the terminal vertex parts in the bubble diagram, which are approximated in our calculation, but do not appear in the measured susceptibility (\ref{Susc2}). Additionally, the form factor in (\ref{Susc2}) is naively neglected as in previous treatments of different models of the bound state\cite{Riseborough2000}. Nonetheless, the qualitative momentum dependence of $A_{\bf q}$ is very similar to the experimental observation.

\section{Conclusions}\label{secConclusions}

The neutron resonance peak observed in SmB$_6$ is well described as a bound-state exciton arising from strong Coulomb repulsion of ``heavy'' $f$ electrons. The existence of this coherent collective excitation in a three-dimensional insulator is a direct evidence of strong correlations, which undoubtedly characterize the ground state as well, and could lead to correlated surface states protected by a non-trivial band topology \cite{Nikolic2014b}. The $f$ electrons are, however, not localized in mixed-valence Kondo insulators such as SmB$_6$; the intrinsic $f$ band is split by hybridization with the $d$ band, and the Fermi level lies in the resulting bandgap. Quasiparticle excitations are thus well defined as ``heavily dressed'' electrons and holes that originate from the hybridized samarium's $f$ and $d$ orbitals. The low-energy quasiparticle spectrum discerned from the neutron scattering experiment is significantly renormalized by Coulomb interactions, and has a band-insulating structure with a non-trivial topology \cite{Fuhrman2014}.

Our theoretical calculation of the collective mode spectrum starts with the Anderson impurity Hamiltonian that can properly handle the mobile $f$ electrons. This model is further adapted for the description of a non-trivial band topology \cite{Dzero2010, Dzero2012}, and then treated to a slave boson approximation in order to make the strong Coulomb interaction regime tractable by perturbation theory. Our subsequent RPA calculation takes into account the interactions between quasiparticles and quasiholes that can directly produce an exciton bound state, and then includes all self-energy corrections to the exciton's dynamics. The quasiparticle spectrum and a few phenomenological parameters that arise from self-consistent renormalizations are estimated directly from the neutron scattering experiment. As a result, the coherent mode spectrum that we calculate matches the neutron-measured dispersion bandwidth and shape with a reasonable quantitative accuracy. The momentum-space distribution of the mode's intensity is matched qualitatively, even though our calculation is incomplete in this regard. The mode's long life-time and coherence is explained by its energy within the particle-hole gap, which eliminates decay channels through energy conservation.

Our method leaves room for a more sophisticated determination of the low-energy quasiparticle spectrum from the neutron data. Since the calculation of the collective mode spectrum can seemingly achieve a reasonable quantitative accuracy, it is plausible that the quasiparticle spectrum could be fitted with more detail than we have done so far. This is left for future work. It will be further helpful to observe and calculate the collective mode spectrum in strong magnetic fields, where momentum-dependent Zeeman splitting could provide a detailed information about the spin-orbit coupling.

\section{Acknowledgements}

We thank Collin Broholm, Tyrel McQueen, Martin Mourigal and Yuan Wan for enlightening discussions.  The work at IQM was supported by the US Department of Energy, office of Basic Energy Sciences, Division of Material Sciences and Engineering under grant DE-FG02-08ER46544.  WTF is grateful for the support of the Gardner Fellowship.

%


%
   \bibliography{SmB6}

\begin{thebibliography}{54}%
\makeatletter
\providecommand \@ifxundefined [1]{%
 \@ifx{#1\undefined}
}%
\providecommand \@ifnum [1]{%
 \ifnum #1\expandafter \@firstoftwo
 \else \expandafter \@secondoftwo
 \fi
}%
\providecommand \@ifx [1]{%
 \ifx #1\expandafter \@firstoftwo
 \else \expandafter \@secondoftwo
 \fi
}%
\providecommand \natexlab [1]{#1}%
\providecommand \enquote  [1]{``#1''}%
\providecommand \bibnamefont  [1]{#1}%
\providecommand \bibfnamefont [1]{#1}%
\providecommand \citenamefont [1]{#1}%
\providecommand \href@noop [0]{\@secondoftwo}%
\providecommand \href [0]{\begingroup \@sanitize@url \@href}%
\providecommand \@href[1]{\@@startlink{#1}\@@href}%
\providecommand \@@href[1]{\endgroup#1\@@endlink}%
\providecommand \@sanitize@url [0]{\catcode `\\12\catcode `\$12\catcode
  `\&12\catcode `\#12\catcode `\^12\catcode `\_12\catcode `\%12\relax}%
\providecommand \@@startlink[1]{}%
\providecommand \@@endlink[0]{}%
\providecommand \url  [0]{\begingroup\@sanitize@url \@url }%
\providecommand \@url [1]{\endgroup\@href {#1}{\urlprefix }}%
\providecommand \urlprefix  [0]{URL }%
\providecommand \Eprint [0]{\href }%
\providecommand \doibase [0]{http://dx.doi.org/}%
\providecommand \selectlanguage [0]{\@gobble}%
\providecommand \bibinfo  [0]{\@secondoftwo}%
\providecommand \bibfield  [0]{\@secondoftwo}%
\providecommand \translation [1]{[#1]}%
\providecommand \BibitemOpen [0]{}%
\providecommand \bibitemStop [0]{}%
\providecommand \bibitemNoStop [0]{.\EOS\space}%
\providecommand \EOS [0]{\spacefactor3000\relax}%
\providecommand \BibitemShut  [1]{\csname bibitem#1\endcsname}%
\let\auto@bib@innerbib\@empty
\bibitem [{\citenamefont {Menth}\ \emph {et~al.}(1969)\citenamefont {Menth},
  \citenamefont {Buehler},\ and\ \citenamefont {Geballe}}]{Menth1969}%
  \BibitemOpen
  \bibfield  {author} {\bibinfo {author} {\bibfnamefont {A.}~\bibnamefont
  {Menth}}, \bibinfo {author} {\bibfnamefont {E.}~\bibnamefont {Buehler}}, \
  and\ \bibinfo {author} {\bibfnamefont {T.~H.}\ \bibnamefont {Geballe}},\
  }\href@noop {} {\bibfield  {journal} {\bibinfo  {journal} {Physical Review
  Letters}\ }\textbf {\bibinfo {volume} {22}},\ \bibinfo {pages} {295}
  (\bibinfo {year} {1969})}\BibitemShut {NoStop}%
\bibitem [{\citenamefont {Nickerson}\ \emph {et~al.}(1971)\citenamefont
  {Nickerson}, \citenamefont {White}, \citenamefont {Lee}, \citenamefont
  {Bachmann}, \citenamefont {Geballe},\ and\ \citenamefont {{G. W.
  Hull}}}]{Nickerson1971}%
  \BibitemOpen
  \bibfield  {author} {\bibinfo {author} {\bibfnamefont {J.~C.}\ \bibnamefont
  {Nickerson}}, \bibinfo {author} {\bibfnamefont {R.~M.}\ \bibnamefont
  {White}}, \bibinfo {author} {\bibfnamefont {K.~N.}\ \bibnamefont {Lee}},
  \bibinfo {author} {\bibfnamefont {R.}~\bibnamefont {Bachmann}}, \bibinfo
  {author} {\bibfnamefont {T.~H.}\ \bibnamefont {Geballe}}, \ and\ \bibinfo
  {author} {\bibfnamefont {J.}~\bibnamefont {{G. W. Hull}}},\ }\href@noop {}
  {\bibfield  {journal} {\bibinfo  {journal} {Physical Review B}\ }\textbf
  {\bibinfo {volume} {3}},\ \bibinfo {pages} {2030} (\bibinfo {year}
  {1971})}\BibitemShut {NoStop}%
\bibitem [{\citenamefont {Hundley}\ \emph {et~al.}(1990)\citenamefont
  {Hundley}, \citenamefont {Canfield}, \citenamefont {Thompson}, \citenamefont
  {Fisk},\ and\ \citenamefont {Lawrence}}]{Hundley1990}%
  \BibitemOpen
  \bibfield  {author} {\bibinfo {author} {\bibfnamefont {M.~F.}\ \bibnamefont
  {Hundley}}, \bibinfo {author} {\bibfnamefont {P.~C.}\ \bibnamefont
  {Canfield}}, \bibinfo {author} {\bibfnamefont {J.~D.}\ \bibnamefont
  {Thompson}}, \bibinfo {author} {\bibfnamefont {Z.}~\bibnamefont {Fisk}}, \
  and\ \bibinfo {author} {\bibfnamefont {J.~M.}\ \bibnamefont {Lawrence}},\
  }\href@noop {} {\bibfield  {journal} {\bibinfo  {journal} {Physical Review
  B}\ }\textbf {\bibinfo {volume} {42}},\ \bibinfo {pages} {6842} (\bibinfo
  {year} {1990})}\BibitemShut {NoStop}%
\bibitem [{\citenamefont {Riseborough}(1992)}]{Riseborough1992}%
  \BibitemOpen
  \bibfield  {author} {\bibinfo {author} {\bibfnamefont {P.~S.}\ \bibnamefont
  {Riseborough}},\ }\href@noop {} {\bibfield  {journal} {\bibinfo  {journal}
  {Physical Review B}\ }\textbf {\bibinfo {volume} {45}},\ \bibinfo {pages}
  {13984} (\bibinfo {year} {1992})}\BibitemShut {NoStop}%
\bibitem [{\citenamefont {Alekseev}\ \emph {et~al.}(1993)\citenamefont
  {Alekseev}, \citenamefont {Mignot}, \citenamefont {Rossat-Mignod},
  \citenamefont {Lazukov},\ and\ \citenamefont {Sadikov}}]{Alekseev1993}%
  \BibitemOpen
  \bibfield  {author} {\bibinfo {author} {\bibfnamefont {P.~A.}\ \bibnamefont
  {Alekseev}}, \bibinfo {author} {\bibfnamefont {J.-M.}\ \bibnamefont
  {Mignot}}, \bibinfo {author} {\bibfnamefont {J.}~\bibnamefont
  {Rossat-Mignod}}, \bibinfo {author} {\bibfnamefont {V.~N.}\ \bibnamefont
  {Lazukov}}, \ and\ \bibinfo {author} {\bibfnamefont {I.~P.}\ \bibnamefont
  {Sadikov}},\ }\href@noop {} {\bibfield  {journal} {\bibinfo  {journal}
  {Physica B}\ }\textbf {\bibinfo {volume} {186--188}},\ \bibinfo {pages} {384}
  (\bibinfo {year} {1993})}\BibitemShut {NoStop}%
\bibitem [{\citenamefont {Nyhus}\ \emph {et~al.}(1995)\citenamefont {Nyhus},
  \citenamefont {Cooper}, \citenamefont {Fisk},\ and\ \citenamefont
  {Sarrao}}]{Nyhus1995}%
  \BibitemOpen
  \bibfield  {author} {\bibinfo {author} {\bibfnamefont {P.}~\bibnamefont
  {Nyhus}}, \bibinfo {author} {\bibfnamefont {S.~L.}\ \bibnamefont {Cooper}},
  \bibinfo {author} {\bibfnamefont {Z.}~\bibnamefont {Fisk}}, \ and\ \bibinfo
  {author} {\bibfnamefont {J.}~\bibnamefont {Sarrao}},\ }\href@noop {}
  {\bibfield  {journal} {\bibinfo  {journal} {Physical Review B}\ }\textbf
  {\bibinfo {volume} {52}},\ \bibinfo {pages} {R14308} (\bibinfo {year}
  {1995})}\BibitemShut {NoStop}%
\bibitem [{\citenamefont {Sera}\ \emph {et~al.}(1996)\citenamefont {Sera},
  \citenamefont {Kobayashi}, \citenamefont {Hiroi}, \citenamefont {Kobayashi},\
  and\ \citenamefont {Kunii}}]{Sera1996}%
  \BibitemOpen
  \bibfield  {author} {\bibinfo {author} {\bibfnamefont {M.}~\bibnamefont
  {Sera}}, \bibinfo {author} {\bibfnamefont {S.}~\bibnamefont {Kobayashi}},
  \bibinfo {author} {\bibfnamefont {M.}~\bibnamefont {Hiroi}}, \bibinfo
  {author} {\bibfnamefont {N.}~\bibnamefont {Kobayashi}}, \ and\ \bibinfo
  {author} {\bibfnamefont {S.}~\bibnamefont {Kunii}},\ }\href@noop {}
  {\bibfield  {journal} {\bibinfo  {journal} {Physical Review B}\ }\textbf
  {\bibinfo {volume} {54}},\ \bibinfo {pages} {R5207} (\bibinfo {year}
  {1996})}\BibitemShut {NoStop}%
\bibitem [{\citenamefont {Okamura}\ \emph {et~al.}(1998)\citenamefont
  {Okamura}, \citenamefont {Kimura}, \citenamefont {Shinozaki}, \citenamefont
  {Nanba}, \citenamefont {Iga}, \citenamefont {Shimizu},\ and\ \citenamefont
  {Takabatake}}]{Okamura1998}%
  \BibitemOpen
  \bibfield  {author} {\bibinfo {author} {\bibfnamefont {H.}~\bibnamefont
  {Okamura}}, \bibinfo {author} {\bibfnamefont {S.}~\bibnamefont {Kimura}},
  \bibinfo {author} {\bibfnamefont {H.}~\bibnamefont {Shinozaki}}, \bibinfo
  {author} {\bibfnamefont {T.}~\bibnamefont {Nanba}}, \bibinfo {author}
  {\bibfnamefont {F.}~\bibnamefont {Iga}}, \bibinfo {author} {\bibfnamefont
  {N.}~\bibnamefont {Shimizu}}, \ and\ \bibinfo {author} {\bibfnamefont
  {T.}~\bibnamefont {Takabatake}},\ }\href@noop {} {\bibfield  {journal}
  {\bibinfo  {journal} {Physical Review B}\ }\textbf {\bibinfo {volume} {58}},\
  \bibinfo {pages} {R7496} (\bibinfo {year} {1998})}\BibitemShut {NoStop}%
\bibitem [{\citenamefont {Bouvet}\ \emph {et~al.}(1998)\citenamefont {Bouvet},
  \citenamefont {Kasuya}, \citenamefont {Bonnet}, \citenamefont {Regnault},
  \citenamefont {Rossat-Mignod}, \citenamefont {Iga}, \citenamefont {Fak},\
  and\ \citenamefont {Severing}}]{Bouvet1998}%
  \BibitemOpen
  \bibfield  {author} {\bibinfo {author} {\bibfnamefont {A.}~\bibnamefont
  {Bouvet}}, \bibinfo {author} {\bibfnamefont {T.}~\bibnamefont {Kasuya}},
  \bibinfo {author} {\bibfnamefont {M.}~\bibnamefont {Bonnet}}, \bibinfo
  {author} {\bibfnamefont {L.~P.}\ \bibnamefont {Regnault}}, \bibinfo {author}
  {\bibfnamefont {J.}~\bibnamefont {Rossat-Mignod}}, \bibinfo {author}
  {\bibfnamefont {F.}~\bibnamefont {Iga}}, \bibinfo {author} {\bibfnamefont
  {B.}~\bibnamefont {Fak}}, \ and\ \bibinfo {author} {\bibfnamefont
  {A.}~\bibnamefont {Severing}},\ }\href@noop {} {\bibfield  {journal}
  {\bibinfo  {journal} {Journal of Physics: Condensed Matter}\ }\textbf
  {\bibinfo {volume} {10}},\ \bibinfo {pages} {5667} (\bibinfo {year}
  {1998})}\BibitemShut {NoStop}%
\bibitem [{\citenamefont {Gorshunov}\ \emph
  {et~al.}(1999{\natexlab{a}})\citenamefont {Gorshunov}, \citenamefont
  {Sluchanko}, \citenamefont {Volkov}, \citenamefont {Dressel}, \citenamefont
  {Knebel}, \citenamefont {Loidl},\ and\ \citenamefont
  {Kunii}}]{Gorshunov1999}%
  \BibitemOpen
  \bibfield  {author} {\bibinfo {author} {\bibfnamefont {B.}~\bibnamefont
  {Gorshunov}}, \bibinfo {author} {\bibfnamefont {N.}~\bibnamefont
  {Sluchanko}}, \bibinfo {author} {\bibfnamefont {A.}~\bibnamefont {Volkov}},
  \bibinfo {author} {\bibfnamefont {M.}~\bibnamefont {Dressel}}, \bibinfo
  {author} {\bibfnamefont {G.}~\bibnamefont {Knebel}}, \bibinfo {author}
  {\bibfnamefont {A.}~\bibnamefont {Loidl}}, \ and\ \bibinfo {author}
  {\bibfnamefont {S.}~\bibnamefont {Kunii}},\ }\href@noop {} {\bibfield
  {journal} {\bibinfo  {journal} {Physical Review B}\ }\textbf {\bibinfo
  {volume} {59}},\ \bibinfo {pages} {1808} (\bibinfo {year}
  {1999}{\natexlab{a}})}\BibitemShut {NoStop}%
\bibitem [{\citenamefont {Riseborough}(2000)}]{Riseborough2000}%
  \BibitemOpen
  \bibfield  {author} {\bibinfo {author} {\bibfnamefont {P.~S.}\ \bibnamefont
  {Riseborough}},\ }\href@noop {} {\bibfield  {journal} {\bibinfo  {journal}
  {Annals of Physics}\ }\textbf {\bibinfo {volume} {9}},\ \bibinfo {pages}
  {813} (\bibinfo {year} {2000})}\BibitemShut {NoStop}%
\bibitem [{\citenamefont {Gorshunov}\ \emph
  {et~al.}(1999{\natexlab{b}})\citenamefont {Gorshunov}, \citenamefont
  {Sluchanko}, \citenamefont {Volkov}, \citenamefont {Dressel}, \citenamefont
  {Knebel}, \citenamefont {Loidl},\ and\ \citenamefont
  {Kunii}}]{PhysRevB.59.1808}%
  \BibitemOpen
  \bibfield  {author} {\bibinfo {author} {\bibfnamefont {B.}~\bibnamefont
  {Gorshunov}}, \bibinfo {author} {\bibfnamefont {N.}~\bibnamefont
  {Sluchanko}}, \bibinfo {author} {\bibfnamefont {A.}~\bibnamefont {Volkov}},
  \bibinfo {author} {\bibfnamefont {M.}~\bibnamefont {Dressel}}, \bibinfo
  {author} {\bibfnamefont {G.}~\bibnamefont {Knebel}}, \bibinfo {author}
  {\bibfnamefont {A.}~\bibnamefont {Loidl}}, \ and\ \bibinfo {author}
  {\bibfnamefont {S.}~\bibnamefont {Kunii}},\ }\href {\doibase
  10.1103/PhysRevB.59.1808} {\bibfield  {journal} {\bibinfo  {journal} {Phys.
  Rev. B}\ }\textbf {\bibinfo {volume} {59}},\ \bibinfo {pages} {1808}
  (\bibinfo {year} {1999}{\natexlab{b}})}\BibitemShut {NoStop}%
\bibitem [{\citenamefont {Hundley}\ \emph {et~al.}(1994)\citenamefont
  {Hundley}, \citenamefont {Thompson}, \citenamefont {Canfield},\ and\
  \citenamefont {Fisk}}]{Hundley1994}%
  \BibitemOpen
  \bibfield  {author} {\bibinfo {author} {\bibfnamefont {M.~F.}\ \bibnamefont
  {Hundley}}, \bibinfo {author} {\bibfnamefont {J.~D.}\ \bibnamefont
  {Thompson}}, \bibinfo {author} {\bibfnamefont {P.~C.}\ \bibnamefont
  {Canfield}}, \ and\ \bibinfo {author} {\bibfnamefont {Z.}~\bibnamefont
  {Fisk}},\ }\href@noop {} {\bibfield  {journal} {\bibinfo  {journal} {Physica
  B}\ }\textbf {\bibinfo {volume} {199-200}},\ \bibinfo {pages} {443} (\bibinfo
  {year} {1994})}\BibitemShut {NoStop}%
\bibitem [{\citenamefont {Kasuya}(1994)}]{Kasuya1994}%
  \BibitemOpen
  \bibfield  {author} {\bibinfo {author} {\bibfnamefont {T.}~\bibnamefont
  {Kasuya}},\ }\href@noop {} {\bibfield  {journal} {\bibinfo  {journal}
  {Europhysics Letters}\ }\textbf {\bibinfo {volume} {26}},\ \bibinfo {pages}
  {277} (\bibinfo {year} {1994})}\BibitemShut {NoStop}%
\bibitem [{\citenamefont {Fuhrman}\ \emph {et~al.}(2014)\citenamefont
  {Fuhrman}, \citenamefont {Leiner}, \citenamefont {Nikoli{\'c}}, \citenamefont
  {Granroth}, \citenamefont {Stone}, \citenamefont {Lumsden}, \citenamefont
  {DeBeer-Schmitt}, \citenamefont {Alekseev}, \citenamefont {Mignot},
  \citenamefont {Koohpayeh}, \citenamefont {Cottingham}, \citenamefont
  {Phelan}, \citenamefont {Schoop}, \citenamefont {McQueen},\ and\
  \citenamefont {Broholm}}]{Fuhrman2014}%
  \BibitemOpen
  \bibfield  {author} {\bibinfo {author} {\bibfnamefont {W.~T.}\ \bibnamefont
  {Fuhrman}}, \bibinfo {author} {\bibfnamefont {J.}~\bibnamefont {Leiner}},
  \bibinfo {author} {\bibfnamefont {P.}~\bibnamefont {Nikoli{\'c}}}, \bibinfo
  {author} {\bibfnamefont {G.~E.}\ \bibnamefont {Granroth}}, \bibinfo {author}
  {\bibfnamefont {M.~B.}\ \bibnamefont {Stone}}, \bibinfo {author}
  {\bibfnamefont {M.~D.}\ \bibnamefont {Lumsden}}, \bibinfo {author}
  {\bibfnamefont {L.}~\bibnamefont {DeBeer-Schmitt}}, \bibinfo {author}
  {\bibfnamefont {P.~A.}\ \bibnamefont {Alekseev}}, \bibinfo {author}
  {\bibfnamefont {J.-M.}\ \bibnamefont {Mignot}}, \bibinfo {author}
  {\bibfnamefont {S.~M.}\ \bibnamefont {Koohpayeh}}, \bibinfo {author}
  {\bibfnamefont {P.}~\bibnamefont {Cottingham}}, \bibinfo {author}
  {\bibfnamefont {W.~A.}\ \bibnamefont {Phelan}}, \bibinfo {author}
  {\bibfnamefont {L.}~\bibnamefont {Schoop}}, \bibinfo {author} {\bibfnamefont
  {T.~M.}\ \bibnamefont {McQueen}}, \ and\ \bibinfo {author} {\bibfnamefont
  {C.}~\bibnamefont {Broholm}},\ }\href@noop {} {\  (\bibinfo {year} {2014})},\
  \bibinfo {note} {arXiv:1407.2647}\BibitemShut {NoStop}%
\bibitem [{\citenamefont {Chazalviel}\ \emph {et~al.}(1976)\citenamefont
  {Chazalviel}, \citenamefont {Campagna}, \citenamefont {Wertheim},\ and\
  \citenamefont {Schmidt}}]{Chazalviel1976}%
  \BibitemOpen
  \bibfield  {author} {\bibinfo {author} {\bibfnamefont {J.~N.}\ \bibnamefont
  {Chazalviel}}, \bibinfo {author} {\bibfnamefont {M.}~\bibnamefont
  {Campagna}}, \bibinfo {author} {\bibfnamefont {G.~K.}\ \bibnamefont
  {Wertheim}}, \ and\ \bibinfo {author} {\bibfnamefont {P.~H.}\ \bibnamefont
  {Schmidt}},\ }\href@noop {} {\bibfield  {journal} {\bibinfo  {journal}
  {Physical Review B}\ }\textbf {\bibinfo {volume} {14}},\ \bibinfo {pages}
  {4586} (\bibinfo {year} {1976})}\BibitemShut {NoStop}%
\bibitem [{\citenamefont {Beaurepaire}\ \emph {et~al.}(1990)\citenamefont
  {Beaurepaire}, \citenamefont {Kappler},\ and\ \citenamefont
  {Krill}}]{Beaurepaire1990}%
  \BibitemOpen
  \bibfield  {author} {\bibinfo {author} {\bibfnamefont {E.}~\bibnamefont
  {Beaurepaire}}, \bibinfo {author} {\bibfnamefont {J.~P.}\ \bibnamefont
  {Kappler}}, \ and\ \bibinfo {author} {\bibfnamefont {G.}~\bibnamefont
  {Krill}},\ }\href@noop {} {\bibfield  {journal} {\bibinfo  {journal}
  {Physical Review B}\ }\textbf {\bibinfo {volume} {41}},\ \bibinfo {pages}
  {6768} (\bibinfo {year} {1990})}\BibitemShut {NoStop}%
\bibitem [{\citenamefont {Dzero}\ \emph {et~al.}(2010)\citenamefont {Dzero},
  \citenamefont {Sun}, \citenamefont {Galitski},\ and\ \citenamefont
  {Coleman}}]{Dzero2010}%
  \BibitemOpen
  \bibfield  {author} {\bibinfo {author} {\bibfnamefont {M.}~\bibnamefont
  {Dzero}}, \bibinfo {author} {\bibfnamefont {K.}~\bibnamefont {Sun}}, \bibinfo
  {author} {\bibfnamefont {V.}~\bibnamefont {Galitski}}, \ and\ \bibinfo
  {author} {\bibfnamefont {P.}~\bibnamefont {Coleman}},\ }\href@noop {}
  {\bibfield  {journal} {\bibinfo  {journal} {Physical Review Letters}\
  }\textbf {\bibinfo {volume} {104}},\ \bibinfo {pages} {106408} (\bibinfo
  {year} {2010})}\BibitemShut {NoStop}%
\bibitem [{\citenamefont {Takimoto}(2011)}]{Takimoto2011}%
  \BibitemOpen
  \bibfield  {author} {\bibinfo {author} {\bibfnamefont {T.}~\bibnamefont
  {Takimoto}},\ }\href@noop {} {\bibfield  {journal} {\bibinfo  {journal}
  {Journal of the Physical Society of Japan}\ }\textbf {\bibinfo {volume}
  {80}},\ \bibinfo {pages} {123710} (\bibinfo {year} {2011})}\BibitemShut
  {NoStop}%
\bibitem [{\citenamefont {Dzero}\ \emph {et~al.}(2012)\citenamefont {Dzero},
  \citenamefont {Sun}, \citenamefont {Coleman},\ and\ \citenamefont
  {Galitski}}]{Dzero2012}%
  \BibitemOpen
  \bibfield  {author} {\bibinfo {author} {\bibfnamefont {M.}~\bibnamefont
  {Dzero}}, \bibinfo {author} {\bibfnamefont {K.}~\bibnamefont {Sun}}, \bibinfo
  {author} {\bibfnamefont {P.}~\bibnamefont {Coleman}}, \ and\ \bibinfo
  {author} {\bibfnamefont {V.}~\bibnamefont {Galitski}},\ }\href@noop {}
  {\bibfield  {journal} {\bibinfo  {journal} {Physical Review B}\ }\textbf
  {\bibinfo {volume} {85}},\ \bibinfo {pages} {045130} (\bibinfo {year}
  {2012})}\BibitemShut {NoStop}%
\bibitem [{\citenamefont {Alexandrov}\ \emph {et~al.}(2013)\citenamefont
  {Alexandrov}, \citenamefont {Dzero},\ and\ \citenamefont
  {Coleman}}]{Alex2013}%
  \BibitemOpen
  \bibfield  {author} {\bibinfo {author} {\bibfnamefont {V.}~\bibnamefont
  {Alexandrov}}, \bibinfo {author} {\bibfnamefont {M.}~\bibnamefont {Dzero}}, \
  and\ \bibinfo {author} {\bibfnamefont {P.}~\bibnamefont {Coleman}},\
  }\href@noop {} {\bibfield  {journal} {\bibinfo  {journal} {Physical Review
  Letters}\ }\textbf {\bibinfo {volume} {111}},\ \bibinfo {pages} {226403}
  (\bibinfo {year} {2013})}\BibitemShut {NoStop}%
\bibitem [{\citenamefont {Dzero}\ and\ \citenamefont
  {Galitski}(2013)}]{Dzero2013}%
  \BibitemOpen
  \bibfield  {author} {\bibinfo {author} {\bibfnamefont {M.}~\bibnamefont
  {Dzero}}\ and\ \bibinfo {author} {\bibfnamefont {V.}~\bibnamefont
  {Galitski}},\ }\href@noop {} {\  (\bibinfo {year} {2013})},\ \bibinfo {note}
  {arXiv:1304.7828}\BibitemShut {NoStop}%
\bibitem [{\citenamefont {Zhang}\ \emph {et~al.}(2013)\citenamefont {Zhang},
  \citenamefont {Butch}, \citenamefont {Syers}, \citenamefont {Ziemak},
  \citenamefont {Greene},\ and\ \citenamefont {Paglione}}]{Zhang2013}%
  \BibitemOpen
  \bibfield  {author} {\bibinfo {author} {\bibfnamefont {X.}~\bibnamefont
  {Zhang}}, \bibinfo {author} {\bibfnamefont {N.~P.}\ \bibnamefont {Butch}},
  \bibinfo {author} {\bibfnamefont {P.}~\bibnamefont {Syers}}, \bibinfo
  {author} {\bibfnamefont {S.}~\bibnamefont {Ziemak}}, \bibinfo {author}
  {\bibfnamefont {R.~L.}\ \bibnamefont {Greene}}, \ and\ \bibinfo {author}
  {\bibfnamefont {J.}~\bibnamefont {Paglione}},\ }\href@noop {} {\bibfield
  {journal} {\bibinfo  {journal} {Physical Review X}\ }\textbf {\bibinfo
  {volume} {3}},\ \bibinfo {pages} {011011} (\bibinfo {year}
  {2013})}\BibitemShut {NoStop}%
\bibitem [{\citenamefont {Wolgast}\ \emph
  {et~al.}(2013{\natexlab{a}})\citenamefont {Wolgast}, \citenamefont {Kurdak},
  \citenamefont {Sun}, \citenamefont {Allen}, \citenamefont {Kim},\ and\
  \citenamefont {Fisk}}]{Wolgast2013}%
  \BibitemOpen
  \bibfield  {author} {\bibinfo {author} {\bibfnamefont {S.}~\bibnamefont
  {Wolgast}}, \bibinfo {author} {\bibfnamefont {C.}~\bibnamefont {Kurdak}},
  \bibinfo {author} {\bibfnamefont {K.}~\bibnamefont {Sun}}, \bibinfo {author}
  {\bibfnamefont {J.~W.}\ \bibnamefont {Allen}}, \bibinfo {author}
  {\bibfnamefont {D.-J.}\ \bibnamefont {Kim}}, \ and\ \bibinfo {author}
  {\bibfnamefont {Z.}~\bibnamefont {Fisk}},\ }\href@noop {} {\bibfield
  {journal} {\bibinfo  {journal} {Physical Review B}\ }\textbf {\bibinfo
  {volume} {88}},\ \bibinfo {pages} {180405(R)} (\bibinfo {year}
  {2013}{\natexlab{a}})}\BibitemShut {NoStop}%
\bibitem [{\citenamefont {Kim}\ \emph {et~al.}(2013)\citenamefont {Kim},
  \citenamefont {Thomas}, \citenamefont {Grant}, \citenamefont {Botimer},
  \citenamefont {Fisk},\ and\ \citenamefont {Xia}}]{Kim2013}%
  \BibitemOpen
  \bibfield  {author} {\bibinfo {author} {\bibfnamefont {D.~J.}\ \bibnamefont
  {Kim}}, \bibinfo {author} {\bibfnamefont {S.}~\bibnamefont {Thomas}},
  \bibinfo {author} {\bibfnamefont {T.}~\bibnamefont {Grant}}, \bibinfo
  {author} {\bibfnamefont {J.}~\bibnamefont {Botimer}}, \bibinfo {author}
  {\bibfnamefont {Z.}~\bibnamefont {Fisk}}, \ and\ \bibinfo {author}
  {\bibfnamefont {J.}~\bibnamefont {Xia}},\ }\href@noop {} {\bibfield
  {journal} {\bibinfo  {journal} {Scientific Reports}\ }\textbf {\bibinfo
  {volume} {3}},\ \bibinfo {pages} {3150} (\bibinfo {year} {2013})}\BibitemShut
  {NoStop}%
\bibitem [{\citenamefont {Kim}\ \emph {et~al.}(2014)\citenamefont {Kim},
  \citenamefont {Xia},\ and\ \citenamefont {Fisk}}]{Kim2013a}%
  \BibitemOpen
  \bibfield  {author} {\bibinfo {author} {\bibfnamefont {D.~J.}\ \bibnamefont
  {Kim}}, \bibinfo {author} {\bibfnamefont {J.}~\bibnamefont {Xia}}, \ and\
  \bibinfo {author} {\bibfnamefont {Z.}~\bibnamefont {Fisk}},\ }\href@noop {}
  {\bibfield  {journal} {\bibinfo  {journal} {Nature Materials}\ }\textbf
  {\bibinfo {volume} {13}},\ \bibinfo {pages} {466} (\bibinfo {year} {2014})},\
  \bibinfo {note} {arXiv:1307.0448}\BibitemShut {NoStop}%
\bibitem [{\citenamefont {Thomas}\ \emph {et~al.}(2013)\citenamefont {Thomas},
  \citenamefont {Kim}, \citenamefont {Chung}, \citenamefont {Grant},
  \citenamefont {Fisk},\ and\ \citenamefont {Xia}}]{Thomas2013}%
  \BibitemOpen
  \bibfield  {author} {\bibinfo {author} {\bibfnamefont {S.}~\bibnamefont
  {Thomas}}, \bibinfo {author} {\bibfnamefont {D.~J.}\ \bibnamefont {Kim}},
  \bibinfo {author} {\bibfnamefont {S.~B.}\ \bibnamefont {Chung}}, \bibinfo
  {author} {\bibfnamefont {T.}~\bibnamefont {Grant}}, \bibinfo {author}
  {\bibfnamefont {Z.}~\bibnamefont {Fisk}}, \ and\ \bibinfo {author}
  {\bibfnamefont {J.}~\bibnamefont {Xia}},\ }\href@noop {} {\  (\bibinfo {year}
  {2013})},\ \bibinfo {note} {arXiv:1307.4133}\BibitemShut {NoStop}%
\bibitem [{\citenamefont {Phelan}\ \emph {et~al.}(2014)\citenamefont {Phelan},
  \citenamefont {Koohpayeh}, \citenamefont {Cottingham}, \citenamefont
  {Freeland}, \citenamefont {Leiner}, \citenamefont {Broholm},\ and\
  \citenamefont {McQueen}}]{Phelan2014}%
  \BibitemOpen
  \bibfield  {author} {\bibinfo {author} {\bibfnamefont {W.~A.}\ \bibnamefont
  {Phelan}}, \bibinfo {author} {\bibfnamefont {S.~M.}\ \bibnamefont
  {Koohpayeh}}, \bibinfo {author} {\bibfnamefont {P.}~\bibnamefont
  {Cottingham}}, \bibinfo {author} {\bibfnamefont {J.~W.}\ \bibnamefont
  {Freeland}}, \bibinfo {author} {\bibfnamefont {J.~C.}\ \bibnamefont
  {Leiner}}, \bibinfo {author} {\bibfnamefont {C.~L.}\ \bibnamefont {Broholm}},
  \ and\ \bibinfo {author} {\bibfnamefont {T.~M.}\ \bibnamefont {McQueen}},\
  }\href {\doibase 10.1103/PhysRevX.4.031012} {\bibfield  {journal} {\bibinfo
  {journal} {Phys. Rev. X}\ }\textbf {\bibinfo {volume} {4}},\ \bibinfo {pages}
  {031012} (\bibinfo {year} {2014})}\BibitemShut {NoStop}%
\bibitem [{\citenamefont {Li}\ \emph {et~al.}(2013)\citenamefont {Li},
  \citenamefont {Xiang}, \citenamefont {Yu}, \citenamefont {Asaba},
  \citenamefont {Lawson}, \citenamefont {Cai}, \citenamefont {Tinsman},
  \citenamefont {Berkley}, \citenamefont {Wolgast}, \citenamefont {Eo},
  \citenamefont {Kim}, \citenamefont {Kurdak}, \citenamefont {Allen},
  \citenamefont {Sun}, \citenamefont {Chen}, \citenamefont {Wang},
  \citenamefont {Fisk},\ and\ \citenamefont {Li}}]{Xiang2013}%
  \BibitemOpen
  \bibfield  {author} {\bibinfo {author} {\bibfnamefont {G.}~\bibnamefont
  {Li}}, \bibinfo {author} {\bibfnamefont {Z.}~\bibnamefont {Xiang}}, \bibinfo
  {author} {\bibfnamefont {F.}~\bibnamefont {Yu}}, \bibinfo {author}
  {\bibfnamefont {T.}~\bibnamefont {Asaba}}, \bibinfo {author} {\bibfnamefont
  {B.}~\bibnamefont {Lawson}}, \bibinfo {author} {\bibfnamefont
  {P.}~\bibnamefont {Cai}}, \bibinfo {author} {\bibfnamefont {C.}~\bibnamefont
  {Tinsman}}, \bibinfo {author} {\bibfnamefont {A.}~\bibnamefont {Berkley}},
  \bibinfo {author} {\bibfnamefont {S.}~\bibnamefont {Wolgast}}, \bibinfo
  {author} {\bibfnamefont {Y.~S.}\ \bibnamefont {Eo}}, \bibinfo {author}
  {\bibfnamefont {D.-J.}\ \bibnamefont {Kim}}, \bibinfo {author} {\bibfnamefont
  {C.}~\bibnamefont {Kurdak}}, \bibinfo {author} {\bibfnamefont {J.~W.}\
  \bibnamefont {Allen}}, \bibinfo {author} {\bibfnamefont {K.}~\bibnamefont
  {Sun}}, \bibinfo {author} {\bibfnamefont {X.~H.}\ \bibnamefont {Chen}},
  \bibinfo {author} {\bibfnamefont {Y.~Y.}\ \bibnamefont {Wang}}, \bibinfo
  {author} {\bibfnamefont {Z.}~\bibnamefont {Fisk}}, \ and\ \bibinfo {author}
  {\bibfnamefont {L.}~\bibnamefont {Li}},\ }\href@noop {} {\  (\bibinfo {year}
  {2013})},\ \bibinfo {note} {arXiv:1306.5221}\BibitemShut {NoStop}%
\bibitem [{\citenamefont {R{\"o}{\ss}ler}\ \emph {et~al.}(2013)\citenamefont
  {R{\"o}{\ss}ler}, \citenamefont {Jang}, \citenamefont {Kim}, \citenamefont
  {Tjeng}, \citenamefont {Fisk}, \citenamefont {Steglich},\ and\ \citenamefont
  {Wirth}}]{Rossler2013}%
  \BibitemOpen
  \bibfield  {author} {\bibinfo {author} {\bibfnamefont {S.}~\bibnamefont
  {R{\"o}{\ss}ler}}, \bibinfo {author} {\bibfnamefont {T.-H.}\ \bibnamefont
  {Jang}}, \bibinfo {author} {\bibfnamefont {D.~J.}\ \bibnamefont {Kim}},
  \bibinfo {author} {\bibfnamefont {L.~H.}\ \bibnamefont {Tjeng}}, \bibinfo
  {author} {\bibfnamefont {Z.}~\bibnamefont {Fisk}}, \bibinfo {author}
  {\bibfnamefont {F.}~\bibnamefont {Steglich}}, \ and\ \bibinfo {author}
  {\bibfnamefont {S.}~\bibnamefont {Wirth}},\ }\href@noop {} {\  (\bibinfo
  {year} {2013})}\BibitemShut {NoStop}%
\bibitem [{\citenamefont {Denlinger}\ \emph {et~al.}(2013)\citenamefont
  {Denlinger}, \citenamefont {Allen}, \citenamefont {Kang}, \citenamefont
  {Sun}, \citenamefont {Kim}, \citenamefont {Shim}, \citenamefont {Min},
  \citenamefont {Kim},\ and\ \citenamefont {Fisk}}]{Denlinger2013}%
  \BibitemOpen
  \bibfield  {author} {\bibinfo {author} {\bibfnamefont {J.~D.}\ \bibnamefont
  {Denlinger}}, \bibinfo {author} {\bibfnamefont {J.~W.}\ \bibnamefont
  {Allen}}, \bibinfo {author} {\bibfnamefont {J.-S.}\ \bibnamefont {Kang}},
  \bibinfo {author} {\bibfnamefont {K.}~\bibnamefont {Sun}}, \bibinfo {author}
  {\bibfnamefont {J.-W.}\ \bibnamefont {Kim}}, \bibinfo {author} {\bibfnamefont
  {J.~H.}\ \bibnamefont {Shim}}, \bibinfo {author} {\bibfnamefont {B.~I.}\
  \bibnamefont {Min}}, \bibinfo {author} {\bibfnamefont {D.-J.}\ \bibnamefont
  {Kim}}, \ and\ \bibinfo {author} {\bibfnamefont {Z.}~\bibnamefont {Fisk}},\
  }\href@noop {} {\  (\bibinfo {year} {2013})},\ \bibinfo {note}
  {arXiv:1312.6637}\BibitemShut {NoStop}%
\bibitem [{\citenamefont {Neupane}\ \emph {et~al.}(2013)\citenamefont
  {Neupane}, \citenamefont {Alidoust}, \citenamefont {Xu}, \citenamefont
  {Kondo}, \citenamefont {Ishida}, \citenamefont {Kim}, \citenamefont {Liu},
  \citenamefont {Belopolski}, \citenamefont {Jo}, \citenamefont {Chang},
  \citenamefont {Jeng}, \citenamefont {Durakiewicz}, \citenamefont {Balicas},
  \citenamefont {Lin}, \citenamefont {Bansil}, \citenamefont {Shin},
  \citenamefont {Fisk},\ and\ \citenamefont {Hasan}}]{Neupane2013}%
  \BibitemOpen
  \bibfield  {author} {\bibinfo {author} {\bibfnamefont {M.}~\bibnamefont
  {Neupane}}, \bibinfo {author} {\bibfnamefont {N.}~\bibnamefont {Alidoust}},
  \bibinfo {author} {\bibfnamefont {S.}~\bibnamefont {Xu}}, \bibinfo {author}
  {\bibfnamefont {T.}~\bibnamefont {Kondo}}, \bibinfo {author} {\bibfnamefont
  {Y.}~\bibnamefont {Ishida}}, \bibinfo {author} {\bibfnamefont {D.-J.}\
  \bibnamefont {Kim}}, \bibinfo {author} {\bibfnamefont {C.}~\bibnamefont
  {Liu}}, \bibinfo {author} {\bibfnamefont {I.}~\bibnamefont {Belopolski}},
  \bibinfo {author} {\bibfnamefont {Y.}~\bibnamefont {Jo}}, \bibinfo {author}
  {\bibfnamefont {T.-R.}\ \bibnamefont {Chang}}, \bibinfo {author}
  {\bibfnamefont {H.-T.}\ \bibnamefont {Jeng}}, \bibinfo {author}
  {\bibfnamefont {T.}~\bibnamefont {Durakiewicz}}, \bibinfo {author}
  {\bibfnamefont {L.}~\bibnamefont {Balicas}}, \bibinfo {author} {\bibfnamefont
  {H.}~\bibnamefont {Lin}}, \bibinfo {author} {\bibfnamefont {A.}~\bibnamefont
  {Bansil}}, \bibinfo {author} {\bibfnamefont {S.}~\bibnamefont {Shin}},
  \bibinfo {author} {\bibfnamefont {Z.}~\bibnamefont {Fisk}}, \ and\ \bibinfo
  {author} {\bibfnamefont {M.~Z.}\ \bibnamefont {Hasan}},\ }\href@noop {}
  {\bibfield  {journal} {\bibinfo  {journal} {Nature Communications}\ }\textbf
  {\bibinfo {volume} {4}},\ \bibinfo {pages} {2991} (\bibinfo {year}
  {2013})}\BibitemShut {NoStop}%
\bibitem [{\citenamefont {Jiang}\ \emph {et~al.}(2013)\citenamefont {Jiang},
  \citenamefont {Li}, \citenamefont {Zhang}, \citenamefont {Sun}, \citenamefont
  {Chen}, \citenamefont {Ye}, \citenamefont {Xu}, \citenamefont {Ge},
  \citenamefont {Tan}, \citenamefont {Niu}, \citenamefont {Xia}, \citenamefont
  {Xie}, \citenamefont {Li}, \citenamefont {Chen}, \citenamefont {Wen},\ and\
  \citenamefont {Feng}}]{Jiang2013}%
  \BibitemOpen
  \bibfield  {author} {\bibinfo {author} {\bibfnamefont {J.}~\bibnamefont
  {Jiang}}, \bibinfo {author} {\bibfnamefont {S.}~\bibnamefont {Li}}, \bibinfo
  {author} {\bibfnamefont {T.}~\bibnamefont {Zhang}}, \bibinfo {author}
  {\bibfnamefont {Z.}~\bibnamefont {Sun}}, \bibinfo {author} {\bibfnamefont
  {F.}~\bibnamefont {Chen}}, \bibinfo {author} {\bibfnamefont {Z.~R.}\
  \bibnamefont {Ye}}, \bibinfo {author} {\bibfnamefont {M.}~\bibnamefont {Xu}},
  \bibinfo {author} {\bibfnamefont {Q.~Q.}\ \bibnamefont {Ge}}, \bibinfo
  {author} {\bibfnamefont {S.~Y.}\ \bibnamefont {Tan}}, \bibinfo {author}
  {\bibfnamefont {X.~H.}\ \bibnamefont {Niu}}, \bibinfo {author} {\bibfnamefont
  {M.}~\bibnamefont {Xia}}, \bibinfo {author} {\bibfnamefont {B.~P.}\
  \bibnamefont {Xie}}, \bibinfo {author} {\bibfnamefont {Y.~F.}\ \bibnamefont
  {Li}}, \bibinfo {author} {\bibfnamefont {X.~H.}\ \bibnamefont {Chen}},
  \bibinfo {author} {\bibfnamefont {H.~H.}\ \bibnamefont {Wen}}, \ and\
  \bibinfo {author} {\bibfnamefont {D.~L.}\ \bibnamefont {Feng}},\ }\href@noop
  {} {\bibfield  {journal} {\bibinfo  {journal} {Nature Communications}\
  }\textbf {\bibinfo {volume} {4}},\ \bibinfo {pages} {3010} (\bibinfo {year}
  {2013})}\BibitemShut {NoStop}%
\bibitem [{\citenamefont {Xu}\ \emph {et~al.}(2013)\citenamefont {Xu},
  \citenamefont {Shi}, \citenamefont {Biswas}, \citenamefont {Matt},
  \citenamefont {Dhaka}, \citenamefont {Huang}, \citenamefont {Plumb},
  \citenamefont {Radovi\ifmmode~\acute{c}\else \'{c}\fi{}}, \citenamefont
  {Dil}, \citenamefont {Pomjakushina}, \citenamefont {Conder}, \citenamefont
  {Amato}, \citenamefont {Salman}, \citenamefont {Paul}, \citenamefont {Mesot},
  \citenamefont {Ding},\ and\ \citenamefont {Shi}}]{PhysRevB.88.121102}%
  \BibitemOpen
  \bibfield  {author} {\bibinfo {author} {\bibfnamefont {N.}~\bibnamefont
  {Xu}}, \bibinfo {author} {\bibfnamefont {X.}~\bibnamefont {Shi}}, \bibinfo
  {author} {\bibfnamefont {P.~K.}\ \bibnamefont {Biswas}}, \bibinfo {author}
  {\bibfnamefont {C.~E.}\ \bibnamefont {Matt}}, \bibinfo {author}
  {\bibfnamefont {R.~S.}\ \bibnamefont {Dhaka}}, \bibinfo {author}
  {\bibfnamefont {Y.}~\bibnamefont {Huang}}, \bibinfo {author} {\bibfnamefont
  {N.~C.}\ \bibnamefont {Plumb}}, \bibinfo {author} {\bibfnamefont
  {M.}~\bibnamefont {Radovi\ifmmode~\acute{c}\else \'{c}\fi{}}}, \bibinfo
  {author} {\bibfnamefont {J.~H.}\ \bibnamefont {Dil}}, \bibinfo {author}
  {\bibfnamefont {E.}~\bibnamefont {Pomjakushina}}, \bibinfo {author}
  {\bibfnamefont {K.}~\bibnamefont {Conder}}, \bibinfo {author} {\bibfnamefont
  {A.}~\bibnamefont {Amato}}, \bibinfo {author} {\bibfnamefont
  {Z.}~\bibnamefont {Salman}}, \bibinfo {author} {\bibfnamefont {D.~M.}\
  \bibnamefont {Paul}}, \bibinfo {author} {\bibfnamefont {J.}~\bibnamefont
  {Mesot}}, \bibinfo {author} {\bibfnamefont {H.}~\bibnamefont {Ding}}, \ and\
  \bibinfo {author} {\bibfnamefont {M.}~\bibnamefont {Shi}},\ }\href {\doibase
  10.1103/PhysRevB.88.121102} {\bibfield  {journal} {\bibinfo  {journal} {Phys.
  Rev. B}\ }\textbf {\bibinfo {volume} {88}},\ \bibinfo {pages} {121102}
  (\bibinfo {year} {2013})}\BibitemShut {NoStop}%
\bibitem [{\citenamefont {Wolgast}\ \emph
  {et~al.}(2013{\natexlab{b}})\citenamefont {Wolgast}, \citenamefont {Kurdak},
  \citenamefont {Sun}, \citenamefont {Allen}, \citenamefont {Kim},\ and\
  \citenamefont {Fisk}}]{PhysRevB.88.180405}%
  \BibitemOpen
  \bibfield  {author} {\bibinfo {author} {\bibfnamefont {S.}~\bibnamefont
  {Wolgast}}, \bibinfo {author} {\bibfnamefont {i.~m. c. b. u. i. e. i.~f.}\
  \bibnamefont {Kurdak}}, \bibinfo {author} {\bibfnamefont {K.}~\bibnamefont
  {Sun}}, \bibinfo {author} {\bibfnamefont {J.~W.}\ \bibnamefont {Allen}},
  \bibinfo {author} {\bibfnamefont {D.-J.}\ \bibnamefont {Kim}}, \ and\
  \bibinfo {author} {\bibfnamefont {Z.}~\bibnamefont {Fisk}},\ }\href {\doibase
  10.1103/PhysRevB.88.180405} {\bibfield  {journal} {\bibinfo  {journal} {Phys.
  Rev. B}\ }\textbf {\bibinfo {volume} {88}},\ \bibinfo {pages} {180405}
  (\bibinfo {year} {2013}{\natexlab{b}})}\BibitemShut {NoStop}%
\bibitem [{\citenamefont {Zhang}\ \emph {et~al.}(2009)\citenamefont {Zhang},
  \citenamefont {Liu}, \citenamefont {Qi}, \citenamefont {Dai}, \citenamefont
  {Fang},\ and\ \citenamefont {Zhang}}]{Zhang-2009zr}%
  \BibitemOpen
  \bibfield  {author} {\bibinfo {author} {\bibfnamefont {H.}~\bibnamefont
  {Zhang}}, \bibinfo {author} {\bibfnamefont {C.-X.}\ \bibnamefont {Liu}},
  \bibinfo {author} {\bibfnamefont {X.-L.}\ \bibnamefont {Qi}}, \bibinfo
  {author} {\bibfnamefont {X.}~\bibnamefont {Dai}}, \bibinfo {author}
  {\bibfnamefont {Z.}~\bibnamefont {Fang}}, \ and\ \bibinfo {author}
  {\bibfnamefont {S.-C.}\ \bibnamefont {Zhang}},\ }\href {\doibase
  10.1038/nphys1270} {\bibfield  {journal} {\bibinfo  {journal} {Nat. Phys.}\
  }\textbf {\bibinfo {volume} {5}},\ \bibinfo {pages} {438} (\bibinfo {year}
  {2009})}\BibitemShut {NoStop}%
\bibitem [{\citenamefont {Xia}\ \emph {et~al.}(2009)\citenamefont {Xia},
  \citenamefont {Qian}, \citenamefont {Hsieh}, \citenamefont {Wray},
  \citenamefont {Pal}, \citenamefont {Lin}, \citenamefont {Bansil},
  \citenamefont {Grauer}, \citenamefont {Hor}, \citenamefont {Cava},\ and\
  \citenamefont {Hasan}}]{Xia:2009}%
  \BibitemOpen
  \bibfield  {author} {\bibinfo {author} {\bibfnamefont {Y.}~\bibnamefont
  {Xia}}, \bibinfo {author} {\bibfnamefont {D.}~\bibnamefont {Qian}}, \bibinfo
  {author} {\bibfnamefont {D.}~\bibnamefont {Hsieh}}, \bibinfo {author}
  {\bibfnamefont {L.}~\bibnamefont {Wray}}, \bibinfo {author} {\bibfnamefont
  {A.}~\bibnamefont {Pal}}, \bibinfo {author} {\bibfnamefont {H.}~\bibnamefont
  {Lin}}, \bibinfo {author} {\bibfnamefont {A.}~\bibnamefont {Bansil}},
  \bibinfo {author} {\bibfnamefont {D.}~\bibnamefont {Grauer}}, \bibinfo
  {author} {\bibfnamefont {Y.~S.}\ \bibnamefont {Hor}}, \bibinfo {author}
  {\bibfnamefont {R.~J.}\ \bibnamefont {Cava}}, \ and\ \bibinfo {author}
  {\bibfnamefont {M.~Z.}\ \bibnamefont {Hasan}},\ }\href {\doibase
  10.1038/nphys1274} {\bibfield  {journal} {\bibinfo  {journal} {Nature
  Physics}\ }\textbf {\bibinfo {volume} {5}},\ \bibinfo {pages} {398} (\bibinfo
  {year} {2009})}\BibitemShut {NoStop}%
\bibitem [{\citenamefont {Hsieh}\ \emph {et~al.}(2008)\citenamefont {Hsieh},
  \citenamefont {Qian}, \citenamefont {Wray}, \citenamefont {Xia},
  \citenamefont {Hor}, \citenamefont {Cava},\ and\ \citenamefont
  {Hasan}}]{Hsieh:2008}%
  \BibitemOpen
  \bibfield  {author} {\bibinfo {author} {\bibfnamefont {D.}~\bibnamefont
  {Hsieh}}, \bibinfo {author} {\bibfnamefont {D.}~\bibnamefont {Qian}},
  \bibinfo {author} {\bibfnamefont {L.}~\bibnamefont {Wray}}, \bibinfo {author}
  {\bibfnamefont {Y.}~\bibnamefont {Xia}}, \bibinfo {author} {\bibfnamefont
  {Y.~S.}\ \bibnamefont {Hor}}, \bibinfo {author} {\bibfnamefont {R.~J.}\
  \bibnamefont {Cava}}, \ and\ \bibinfo {author} {\bibfnamefont {M.~Z.}\
  \bibnamefont {Hasan}},\ }\href {\doibase 10.1038/nature06843} {\bibfield
  {journal} {\bibinfo  {journal} {Nature}\ }\textbf {\bibinfo {volume} {452}},\
  \bibinfo {pages} {970} (\bibinfo {year} {2008})}\BibitemShut {NoStop}%
\bibitem [{\citenamefont {Roy}\ \emph {et~al.}(2014)\citenamefont {Roy},
  \citenamefont {Sau}, \citenamefont {Dzero},\ and\ \citenamefont
  {Galitski}}]{Roy2014}%
  \BibitemOpen
  \bibfield  {author} {\bibinfo {author} {\bibfnamefont {B.}~\bibnamefont
  {Roy}}, \bibinfo {author} {\bibfnamefont {J.~D.}\ \bibnamefont {Sau}},
  \bibinfo {author} {\bibfnamefont {M.}~\bibnamefont {Dzero}}, \ and\ \bibinfo
  {author} {\bibfnamefont {V.}~\bibnamefont {Galitski}},\ }\href@noop {} {\
  (\bibinfo {year} {2014})},\ \bibinfo {note} {arXiv:1405.5526}\BibitemShut
  {NoStop}%
\bibitem [{\citenamefont {Nikoli{\'c}}(2014)}]{Nikolic2014b}%
  \BibitemOpen
  \bibfield  {author} {\bibinfo {author} {\bibfnamefont {P.}~\bibnamefont
  {Nikoli{\'c}}},\ }\href@noop {} {\  (\bibinfo {year} {2014})},\ \bibinfo
  {note} {arXiv:1407.4482}\BibitemShut {NoStop}%
\bibitem [{SmB()}]{SmB6LatticeDynamics}%
  \BibitemOpen
  \href@noop {} {\ }\BibitemShut {NoStop}%
\bibitem [{\citenamefont {Miyazaki}\ \emph {et~al.}(2012)\citenamefont
  {Miyazaki}, \citenamefont {Hajiri}, \citenamefont {Ito}, \citenamefont
  {Kunii},\ and\ \citenamefont {ichi Kimura}}]{Miyazaki2012}%
  \BibitemOpen
  \bibfield  {author} {\bibinfo {author} {\bibfnamefont {H.}~\bibnamefont
  {Miyazaki}}, \bibinfo {author} {\bibfnamefont {T.}~\bibnamefont {Hajiri}},
  \bibinfo {author} {\bibfnamefont {T.}~\bibnamefont {Ito}}, \bibinfo {author}
  {\bibfnamefont {S.}~\bibnamefont {Kunii}}, \ and\ \bibinfo {author}
  {\bibfnamefont {S.}~\bibnamefont {ichi Kimura}},\ }\href@noop {} {\bibfield
  {journal} {\bibinfo  {journal} {Physical Review B}\ }\textbf {\bibinfo
  {volume} {86}},\ \bibinfo {pages} {075105} (\bibinfo {year}
  {2012})}\BibitemShut {NoStop}%
\bibitem [{\citenamefont {Kang}\ \emph {et~al.}(2013)\citenamefont {Kang},
  \citenamefont {Kim}, \citenamefont {Kim}, \citenamefont {Kang}, \citenamefont
  {Denlinger},\ and\ \citenamefont {Min}}]{Kang2013}%
  \BibitemOpen
  \bibfield  {author} {\bibinfo {author} {\bibfnamefont {C.-J.}\ \bibnamefont
  {Kang}}, \bibinfo {author} {\bibfnamefont {J.}~\bibnamefont {Kim}}, \bibinfo
  {author} {\bibfnamefont {K.}~\bibnamefont {Kim}}, \bibinfo {author}
  {\bibfnamefont {J.-S.}\ \bibnamefont {Kang}}, \bibinfo {author}
  {\bibfnamefont {J.~D.}\ \bibnamefont {Denlinger}}, \ and\ \bibinfo {author}
  {\bibfnamefont {B.~I.}\ \bibnamefont {Min}},\ }\href@noop {} {\  (\bibinfo
  {year} {2013})}\BibitemShut {NoStop}%
\bibitem [{\citenamefont {Lu}\ \emph {et~al.}(2013)\citenamefont {Lu},
  \citenamefont {Zhao}, \citenamefont {Weng}, \citenamefont {Fang},\ and\
  \citenamefont {Dai}}]{Lu2013b}%
  \BibitemOpen
  \bibfield  {author} {\bibinfo {author} {\bibfnamefont {F.}~\bibnamefont
  {Lu}}, \bibinfo {author} {\bibfnamefont {J.}~\bibnamefont {Zhao}}, \bibinfo
  {author} {\bibfnamefont {H.}~\bibnamefont {Weng}}, \bibinfo {author}
  {\bibfnamefont {Z.}~\bibnamefont {Fang}}, \ and\ \bibinfo {author}
  {\bibfnamefont {X.}~\bibnamefont {Dai}},\ }\href@noop {} {\bibfield
  {journal} {\bibinfo  {journal} {Physical Review Letters}\ }\textbf {\bibinfo
  {volume} {110}},\ \bibinfo {pages} {096401} (\bibinfo {year}
  {2013})}\BibitemShut {NoStop}%
\bibitem [{\citenamefont {Werner}\ and\ \citenamefont
  {Assaad}(2013)}]{Werner2013}%
  \BibitemOpen
  \bibfield  {author} {\bibinfo {author} {\bibfnamefont {J.}~\bibnamefont
  {Werner}}\ and\ \bibinfo {author} {\bibfnamefont {F.~F.}\ \bibnamefont
  {Assaad}},\ }\href@noop {} {\bibfield  {journal} {\bibinfo  {journal}
  {Physical Review B}\ }\textbf {\bibinfo {volume} {88}},\ \bibinfo {pages}
  {035113} (\bibinfo {year} {2013})}\BibitemShut {NoStop}%
\bibitem [{\citenamefont {Legner}\ \emph {et~al.}(2014)\citenamefont {Legner},
  \citenamefont {R{\"u}egg},\ and\ \citenamefont {Sigrist}}]{Legner2014}%
  \BibitemOpen
  \bibfield  {author} {\bibinfo {author} {\bibfnamefont {M.}~\bibnamefont
  {Legner}}, \bibinfo {author} {\bibfnamefont {A.}~\bibnamefont {R{\"u}egg}}, \
  and\ \bibinfo {author} {\bibfnamefont {M.}~\bibnamefont {Sigrist}},\
  }\href@noop {} {\bibfield  {journal} {\bibinfo  {journal} {Physical Review
  B}\ }\textbf {\bibinfo {volume} {89}},\ \bibinfo {pages} {085110} (\bibinfo
  {year} {2014})}\BibitemShut {NoStop}%
\bibitem [{\citenamefont {Kikoin}\ and\ \citenamefont
  {Mishchenko}(1995)}]{KikoinIV}%
  \BibitemOpen
  \bibfield  {author} {\bibinfo {author} {\bibfnamefont {K.~A.}\ \bibnamefont
  {Kikoin}}\ and\ \bibinfo {author} {\bibfnamefont {A.~S.}\ \bibnamefont
  {Mishchenko}},\ }\href {http://stacks.iop.org/0953-8984/7/i=2/a=008}
  {\bibfield  {journal} {\bibinfo  {journal} {Journal of Physics: Condensed
  Matter}\ }\textbf {\bibinfo {volume} {7}},\ \bibinfo {pages} {307} (\bibinfo
  {year} {1995})}\BibitemShut {NoStop}%
\bibitem [{\citenamefont {Riseborough}(2001)}]{Riseborough2001}%
  \BibitemOpen
  \bibfield  {author} {\bibinfo {author} {\bibfnamefont {P.~S.}\ \bibnamefont
  {Riseborough}},\ }\href@noop {} {\bibfield  {journal} {\bibinfo  {journal}
  {Journal of Magnetism and Magnetic Materials}\ }\textbf {\bibinfo {volume}
  {226-230}},\ \bibinfo {pages} {127} (\bibinfo {year} {2001})}\BibitemShut
  {NoStop}%
\bibitem [{Note1()}]{Note1}%
  \BibitemOpen
  \bibinfo {note} {Otherwise, one particular linear combination of $d$ orbitals
  hybridizes with one similar linear combination of $f$ orbitals, while the
  other (orthogonal) linear combinations of these orbitals ruin the insulting
  band-structure by crossing and filling the hybridization gap.}\BibitemShut
  {Stop}%
\bibitem [{\citenamefont {Coleman}(1984)}]{PhysRevB.29.3035}%
  \BibitemOpen
  \bibfield  {author} {\bibinfo {author} {\bibfnamefont {P.}~\bibnamefont
  {Coleman}},\ }\href {\doibase 10.1103/PhysRevB.29.3035} {\bibfield  {journal}
  {\bibinfo  {journal} {Phys. Rev. B}\ }\textbf {\bibinfo {volume} {29}},\
  \bibinfo {pages} {3035} (\bibinfo {year} {1984})}\BibitemShut {NoStop}%
\bibitem [{\citenamefont {Squires}(1996)}]{SquiresTNS}%
  \BibitemOpen
  \bibfield  {author} {\bibinfo {author} {\bibfnamefont {G.~L.}\ \bibnamefont
  {Squires}},\ }\href@noop {} {\emph {\bibinfo {title} {Introduction to the
  Theory of Thermal Neutron Scattering}}}\ (\bibinfo  {publisher} {Dover, New
  York},\ \bibinfo {year} {1996})\BibitemShut {NoStop}%
\bibitem [{\citenamefont {{Denlinger}}\ \emph {et~al.}(2013)\citenamefont
  {{Denlinger}}, \citenamefont {{Allen}}, \citenamefont {{Kang}}, \citenamefont
  {{Sun}}, \citenamefont {{Min}}, \citenamefont {{Kim}},\ and\ \citenamefont
  {{Fisk}}}]{SmB6PastAndPresent}%
  \BibitemOpen
  \bibfield  {author} {\bibinfo {author} {\bibfnamefont {J.~D.}\ \bibnamefont
  {{Denlinger}}}, \bibinfo {author} {\bibfnamefont {J.~W.}\ \bibnamefont
  {{Allen}}}, \bibinfo {author} {\bibfnamefont {J.-S.}\ \bibnamefont {{Kang}}},
  \bibinfo {author} {\bibfnamefont {K.}~\bibnamefont {{Sun}}}, \bibinfo
  {author} {\bibfnamefont {B.-I.}\ \bibnamefont {{Min}}}, \bibinfo {author}
  {\bibfnamefont {D.-J.}\ \bibnamefont {{Kim}}}, \ and\ \bibinfo {author}
  {\bibfnamefont {Z.}~\bibnamefont {{Fisk}}},\ }\href@noop {} {\bibfield
  {journal} {\bibinfo  {journal} {ArXiv e-prints}\ } (\bibinfo {year}
  {2013})},\ \Eprint {http://arxiv.org/abs/1312.6636} {arXiv:1312.6636
  [cond-mat.str-el]} \BibitemShut {NoStop}%
\bibitem [{\citenamefont {Wen}(2004)}]{WenQFT2004}%
  \BibitemOpen
  \bibfield  {author} {\bibinfo {author} {\bibfnamefont {X.-G.}\ \bibnamefont
  {Wen}},\ }\href@noop {} {\emph {\bibinfo {title} {{Quantum Field Theory of
  Many-Body Systems}}}}\ (\bibinfo  {publisher} {Oxford University Press},\
  \bibinfo {address} {New York},\ \bibinfo {year} {2004})\BibitemShut {NoStop}%
\bibitem [{Note2()}]{Note2}%
  \BibitemOpen
  \bibinfo {note} {The slave boson method sees the bandgap as being
  proportional to the slave boson amplitude $|B|$. If the temperature is
  raised, $|B|$ decreases and the bandgap shrinks in agreement with
  experimental observations.}\BibitemShut {Stop}%
\end{thebibliography}%
%

\newpage
\onecolumngrid
\appendix


\end{document}